\documentclass{aa}  

\usepackage{graphicx}
\usepackage{txfonts}
\usepackage{gensymb}
\usepackage[utf8]{inputenc}

\usepackage{natbib}

\usepackage{xcolor}
\definecolor{vincent}{rgb}{0.00, 0.00, 0.5}
\definecolor{kostas}{rgb}{1.00, 0.00, 0.0}
\definecolor{myblue}{rgb}{0.00, 0.0, 0.9}
\definecolor{myred}{rgb}{0.90, 0.0, 0.0}
\definecolor{mygreen}{rgb}{0.0, 0.7, 0.0}
\usepackage[breaklinks,  citecolor=myblue, linkcolor=myred, urlcolor=purple, colorlinks=true, debug, baseurl=' ']{hyperref}

\begin{document}

   \title{Polarization power spectra and dust cloud morphology}

   \subtitle{}

   \author{A. Konstantinou
          \inst{1,2}\fnmsep\thanks{akonstantinou@physics.uoc.gr},
          V. Pelgrims
          \inst{1,2}\fnmsep\thanks{pelgrims@physics.uoc.gr},
          F. Fuchs
          \inst{3}
          \and
          K. Tassis
          \inst{1,2}
          }

   \institute{Department of Physics, University of Crete, Voutes, 70013 Heraklion, Greece
         \and
             Institute of Astrophysics, Foundation for Research and Technology-Hellas, Voutes, 70013 Heraklion, Greece
         \and Department of Computer Science, University of Applied Sciences Trier, 54208 Trier, Germany
             }

   \date{Received November 4, 2021; accepted April 19, 2022}

 
  \abstract
   {In the framework of cosmic microwave background polarization studies and the characterization of its Galactic foregrounds, the angular power spectrum analysis of the thermal dust polarization map has led to intriguing evidence of an $E/B$ asymmetry and a positive $TE$ correlation. The interpretation of these observations are the subject of theoretical and simulation-driven studies
   in which the correlation between the density structure of the interstellar medium (ISM) and the magnetic field appear to be a key aspect.
   In this context, and when the magnetized ISM structures are modeled in three dimensions, dust clouds are generally considered to be filamentary structures only, while both filamentary and sheet-like shapes are supported by observational and theoretical evidence.
   
   }
   {In this work, we aim to study the influence of cloud shape and its connection to the local magnetic field, as well as the influence from the viewing angle, on the angular power spectra measured on thermal dust polarization maps; we specifically focus on the dependence of the $E/B$ power asymmetry and $TE$ correlation.
   }
   {
   To that end, we simulate realistic interstellar clouds with both filament-like and sheet-like shapes using the software \textsc{Asterion} which also allows us to generate synthetic maps of thermal dust polarized emission with an area of 400 square degrees. Then, we compute their polarization power spectra in multipole range $\ell \in [100,\,500]$ and focus on the $E/B$ power asymmetry, quantified through the $\mathcal{R}_{EB}$ ratio, and the correlation coefficient $r^{TE}$ between $T$ and $E$ modes. We quantify the dependence of $\mathcal{R}_{EB}$ and $r^{TE}$ values on the offset angle (between the longest cloud axis and local magnetic field lines) and inclination angle (between the line of sight and the magnetic field) for both types of cloud shapes either embedded in a regular magnetic field or coupled to a non-regular field to mimic turbulence.
   }
   {We find that both types of cloud shapes cover the same regions of the ($\mathcal{R}_{EB},\,r^{TE}$) parameter space. The dependence on the inclination and offset angles are similar for both shapes although sheet-like structures generally show larger scatter as compared to filamentary structures. In addition to the known dependence on the offset angle, we find a strong dependence of $\mathcal{R}_{EB}$ and $r^{TE}$ on the inclination angle.
   }
   {The very fact that filament-like and sheet-like structures may lead to polarization power spectra with similar ($\mathcal{R}_{EB},\,r^{TE}$) values complicates their interpretation. We argue that interpreting them solely in terms of filament characteristics is risky and that in future analyses, this degeneracy should be accounted for as well as the connection to the magnetic field geometry.
   Indeed, our results from 400 square degrees maps make it clear that the overall geometrical arrangement of the magnetized ISM surrounding the observer leaves its marks on polarization power spectra.
   
   }

   
\maketitle
%

\section{Introduction}

The study of polarization of the cosmic microwave background (CMB) is very important for the field of cosmology as it is the only foreseeable way to test observationally  whether the Universe has undergone an early period of inflationary expansion.
Indeed, gravitational waves generated during the inflation era would have left unambiguous imprints on CMB polarization. These imprints are the so-called $B$-modes in the polarization power spectra which, although difficult to detect, are in principle observable (\citealt{polnarev}).

Most of the cosmological information can be obtained from the CMB radiation through the analysis of its $T$, $E$, and $B$ angular power spectra (intensity and linear polarization) and their correlation (e.g., \citealt{Kamionkowski}; \citealt{seljak97}; \citealt{hu}) in which the polarization is decomposed into a scalar, curl-free field, the $E$ mode, and a pseudo-scalar, curl field, the $B$ mode.

Galactic dust, through which we observe the CMB, contaminates the cosmological signal as the grain thermal emission is also polarized and contains $B$-mode components as well. Non-spherical dust grains align their short axis with the magnetic field of the Galaxy which is present everywhere. As a result, they emit thermal radiation with their electric vector preferentially aligned with their long axis (e.g., \citealt{andersson}).
The polarized emission from Galactic dust grains, with polarization fraction about 20\% (e.g., \citealt{Ben2004}; \citealt{planck18}), higher than that of the CMB, which is about 5\% (\citealt{bond}; \citealt{seljak_uro}; \citealt{hinshaw}), always dominates the measured signal at frequency higher than $\approx$ 80~GHz (e.g., \citealt{planck16XXX}).
The $B$-mode component of Galactic thermal dust emission can be confused with the primordial signal of inflation (e.g., \citealt{bicep}; \citealt{planck16XXX}; \citealt{planck20XI}). It is therefore essential to understand the emission and absorption properties of interstellar dust so as to remove its contribution to the map of the polarized primordial microwave sky.

In this context, the dust polarization is characterized by its $T$, $E$, and $B$ representations rather than through its representation in Stokes parameters $I$, $Q$, and $U$. $E$ and $B$ are invariant under rotations making the $E$-$B$ decomposition more convenient since it is unnecessary to define a reference frame.

The characterization of the dust polarization auto- and cross-angular power spectra has led to the intriguing evidences, first reported in (\citealt{planck2016XXXVIII}), for an $E/B$ asymmetry ($\mathcal{R}_{EB}\sim2$) and a positive $TE$ correlation ($r^{TE}\sim0.36$) that were not anticipated, at least for sky regions at intermediate and high Galactic latitudes in which a weak signal with random polarization orientations was expected from the dust.
Dedicated analysis of the data has revealed that this -a priori unexpected- signal may likely originate from the observed correlation between the orientation of ridges seen in intensity maps and the sky-projected orientation of the integrated magnetic field (\citealt{planck2016XXXII}; \citealt{Planck2016XXXV}; \citealt{planck2016XXXVIII}; \citealt{rot2019}; \citealt{soler19}).
Such a correlation between ISM clouds and magnetic field was known for some time for molecular clouds (e.g., \cite{goo1990}) and subsequently observed for clouds of the diffuse ISM seen in H$_{\rm{I}}$ emission (\citealt{McC2006}; \citealt{clark14}).
This observed correlation is likely the result of the coupling of interstellar matter and the ambient magnetic field, which is expected in the framework of the magneto-hydrodynamic (MHD) description of ISM and it has been extensively discussed in the literature (e.g., \citealt{hen2013}; \citealt{sol2013}; \citealt{Cal2017}; \citealt{kandel18}; \citealt{kim2019}).

To make progress in the characterization of Galactic foregrounds to the CMB polarization, it is necessary to consider the sky and the ISM structures as three-dimensional (3D) systems projected onto the sky (e.g., \citealt{tp2015}; \citealt{Pel+21}). In this context, dedicated studies have been carried out to estimate and characterize the result of projection effects of ISM structures to the measured polarization power spectra (e.g., \citealt{rot2019}; \citealt{huffenberger}; \citealt{her2021}). In these studies, the shape of ISM structures have been routinely, if not exclusively, assumed to be filamentary. The authors have indeed considered and modeled ISM clouds as being cylinders or prolate (elongated) spheroids. However, to date, there is no consensus that all ISM clouds have a filament-like morphology.
On the contrary, numerical simulations show that turbulent flows tend to stretch and compress the ISM gas into sheet-like and filament-like structures, which appear elongated in column density maps because of projection effects (e.g., \citealt{hen2013}).
This shape ambiguity is not new and has been at the heart of a long lasting debate in the ISM literature (e.g., \citealt{kai2016} vs. \citealt{TritsisTassis2018}). Dense, molecular, self-gravitating clouds generally appear filamentary on the sky, giving rise to the picture of a cosmic web in which clouds fragment to give rise to star-forming regions at its nodes  (e.g., \citealt{mye2009}). However, there are several observational pieces of evidence indicating the existence of clouds with sheet-like morphology that are necessary to explain light echoes of supernovae explosions and pulsar scattering data (e.g., \citealt{wil1972}; \citealt{suntzeff}; \citealt{spyromilio}; \citealt{brisken};  \citealt{yang}). Such sheet-like clouds may result from large scale planar shocks when supernovae induced bubbles interact \citep{Hartmann2001}.
In addition, the very fact that filamentary patterns are not all straight but may show large-scale curvature, as suggested in (\citealt{planck2016XXXVIII}), might also be indicative of sheet-like structures that would, for example, be part of expanding bubbles, and would appear brighter where viewed tangentially (e.g., \citealt{tah2022}).
Hence, while diverse observations of the diffuse ISM -from relatively low-resolution submillimeter observations by Planck to high-resolution infrared observations by Herschel, and of H$_{\rm{I}}$ observation in the radio waves (e.g., \citealt{kal2009}; \citealt{miv2010}; \citealt{And2014}; \citealt{clark14})- reveal its filamentary appearance, it might very well be that it is made of clouds having both filament-like and sheet-like shapes.

The preferred shape of ISM clouds in the diffuse ISM, along with their specific relation to the magnetic field orientation, is expected to depend on the exact balance of several factors: magnetic field strength, gravitational energy, and turbulent energy (e.g., \citealt{Heiles2005}; \citealt{cru2010}; \citealt{sol2013}).
However, it is very challenging to determine observationally the regime corresponding to any given ISM region. The characterization of relative orientations between apparent structures and projected magnetic field is an equally difficult task, with results that may also depend on the details of the specific method used in the analysis (e.g., \citealt{mic2021}).

In this paper, motivated by the fact that the possibility for sheet-like clouds has been disregarded so far in CMB-foreground characterization studies that want to account for 3D effects, we explore the possible effects from ISM-cloud morphology on the characterization of polarization power spectra.

Our main goal is to study how the shape of interstellar clouds affects the angular power spectra of emitted polarized radiation from dust. We set up a simulation-based experiment to investigate the degeneracy that cloud shapes might lead to in the properties of observed polarization power spectra. Our toy-models allow us to quantify the effect of the angle between the structure-major axes and the local magnetic field orientation, as well as the effect of the viewing angle, i.e. the angle that the sight-lines make with the magnetic field lines.

We simulate interstellar clouds with both filamentary and sheet-like structure using the software \textsc{Asterion} that we developed and present in Sect.~\ref{sec:numerics}, where we also produce synthetic polarization maps and compute their angular power spectra. In Sect.~\ref{sec:analysis} we present the results of our exploration of the effect the two angles have on the power asymmetry between $E$ and $B$ modes, and on the correlation between $T$ and $E$ modes. We discuss our results in Sect.~\ref{sec:discussion} and provide our conclusions in Sect.~\ref{sec:conclusion}.


\section{Numerics \& synthetic data}
\label{sec:numerics}
\subsection{\textsc{Asterion}}
\subsubsection{The software}
 \textsc{Asterion} is a scientific tool to simulate the magnetized ISM of our Galaxy in 3D, including dust clouds and the magnetic field that permeates them, to directly visualize the data in 3D and interact with it into an immersed virtual environment.

Ultimately, we hope that \textsc{Asterion} will assist in the reconstruction of the 3D structure of the magnetized ISM of our Galaxy, a long-standing problem in the quest for the origin of the magnetic field in our Galaxy. \textsc{Asterion} relies on real time 3D visualization techniques with virtual reality capabilities that reach a high degree of performance. It allows us to render the details of the magnetized ISM and enables the user to interact with the simulated environment as it is done in video games.
 
\textsc{Asterion} is not meant to be another visualization and ray-tracing tool, with or without radiative transfer implemented, as already existing ones such as \textsc{Polaris} (\citealt{rei2016}).
\textsc{Asterion} enables real-time reprocessing of portions of the visualized data at a higher resolution, allowing sub-grid details to be computed and added to the visualized data in real-time.
\textsc{Asterion} is implemented in the Real-Time-Engine `Unreal Engine 4' (UE4)\footnote{UE4 is a complete suite of development tools that allows for cutting edge visualization and immersive virtual worlds, multi-platform deployment, asset and plugin marketplace, among other features (\url{https://www.unrealengine.com/en-US/unreal}).}.

The core features of creating and visualizing 3D structures use the parallel-computation paradigm and utilize the benefits of modern Graphics Processing Units (GPU). A significant amount of hardware threads (Shaders) performs calculations independently and presents solutions at once, as 3D textures, available for further computation and visualization. 

Within \textsc{Asterion} (currently at the prototype level), the user is immersed in a 3D environment that emulates the Galaxy. The user can fly through the Galactic space and visualize dust density distribution and magnetic field.
For the purpose of this work the user can specify, through a set of parameters, the large-scale dust-density distribution model, the large-scale Galactic magnetic field model and, clouds of dust with non-trivial morphology following some orientation relation with the ambient magnetic field lines. The structures become accessible using interactive controls and volume-rendering techniques. The user can also specify a virtual-telescope (Observer) in the environment and output two-dimensional (2D) projections (polarization and column density maps) with given angular size. These polarization maps, intended to simulate observations of the polarized thermal dust emission in the sub-millimeter, encode information on the density and the magnetic field properties of the observed portion of the Galaxy. These outputs can then be analyzed externally as if they were actual observations.
To allow for statistical analyses and to make \textsc{Asterion} produce massive amounts of selected outputs, input specifications in the form of setup files (CSV format) can be passed into \textsc{Asterion}, which runs over each and produces the corresponding output maps in an automated way.

In its current implementation, \textsc{Asterion} simulates both the Galaxy at the large scales and dust clouds with higher spatial resolution in a smaller volume. Both volumes are sampled by cubic grids made of $256^3$ voxels. The low-resolution, large-scale, grid is 40 kpc on a side and the high-resolution grid has a size that can be adjusted by the user. Both grids are represented as Volume-Texture on the GPUs and, for each cell, local density and magnetic field as defined by model settings is rendered.

Within \textsc{Asterion}, both the location of the high-resolution observed box (OB) and the observer can be determined by the user who is free to move independently through space. The position of the observer is set in a heliocentric Cartesian coordinate system and the OB location is fixed through specification of the longitude, latitude and distance as seen from the observer. The outputs that are generated by the software and that are of interest in this study (see Sect.~\ref{sec:observables}) are related to the polarized thermal dust emission and correspond to the view of the OB (containing simulated clouds at high-resolution) as seen by the observer. Only the volume spanned by the OB is mapped into the outputs. We make this choice to make it possible, in the future, to simulate large volumes with high-resolution capabilities by the production of mosaic in 3D.

\subsubsection{Polarized thermal dust emission model}
\label{sec:observables}
For a given setup of the simulated magnetized ISM, synthetic observation of the thermal dust polarization and column density can be produced by the software. These come in the form of four maps: $I$, $Q$, $U$ for the three Stokes parameters of the linear polarization of the thermal dust emission, and $K$ for the total column density. These quantities are obtained through integration along the lines of sight through the OB according to the following equations:
\begin{align}
I(\mathbf{n}) \propto & \, 
\int_{0}^{+\infty} dr \, n_{\rm{d}}(r,\mathbf{n}) \, \left\lbrace
1 + \mathfrak{p_0} \left(\frac{2}{3} - \sin^2 \alpha(r,\mathbf{n}) \right)
\right\rbrace
\label{eq:intensity}
\\ 
Q(\mathbf{n}) \propto & \, \mathfrak{p_0}
\int_{0}^{+\infty}{dr \, n_{\rm{d}}(r,\mathbf{n}) \, 
\sin^2 \alpha(r,\mathbf{n}) \, \cos[2\, \psi(r,\mathbf{n})]}
\label{eq:DUSTEMISSION_Q}
\\
U(\mathbf{n}) \propto & \, \mathfrak{p_0}
\int_{0}^{+\infty}{dr \, n_{\rm{d}}(r,\mathbf{n}) \, 
\sin^2 \alpha(r,\mathbf{n}) \, \sin[2\, \psi(r,\mathbf{n})]
}
\label{eq:DUSTEMISSION_U}
\\
K(\mathbf{n}) \propto & \,
\int_{0}^{+\infty} dr \, n_d(r,\mathbf{n})
\label{eq:Columndensity1}
\end{align}
where $r$ is the radial distance from the observer along a given line of sight with direction specified by $\mathbf{n}$.
The different terms in the equation are:
\begin{itemize}
\item $\mathfrak{p_0}$ is a parameter dependent on dust polarization properties (grain cross sections and the degree of alignment with the magnetic field), taking a value of $0.25$ in this work
\footnote{
This is a rather high value compared to the value of $0.2$ used in \cite{planck2015XX}. The choice of this value is motivated by the fact that it leads to average degree of polarization of $\approx 21\%$ in the maps that we generate in Sect.~\ref{sec:setup} when the magnetic field is in the plane of the sky.
We note that the exact value of this parameter does not affect our analysis nor our main conclusion. A purely multiplicative rescaling of $\frak{p}$ affects the scaling of the power spectra but not the parameters $\mathcal{R}_{EB}$ and $r^{TE}$ (\citealt{huffenberger}).
}
\item $n_{\rm{d}}(r,\,\mathbf{n})$ is the 3D dust grain density at position $(r,\,\mathbf{n})$
\item $\alpha(r,\,\mathbf{n})$ is the inclination angle between the magnetic field and the line of sight at $(r,\,\mathbf{n})$
\item $\psi(r,\,\mathbf{n})$ is the local plane-of-sky polarization angle expressed as 
\begin{equation}
\psi(r, \, \mathbf{n}) = \frac{1}{2} \arctan
\left( \frac{-2 \, B_\theta(r, \, \mathbf{n}) \, B_\phi(r, \, \mathbf{n})}
{B_\phi(r, \, \mathbf{n})^2 - B_\theta(r, \, \mathbf{n})^2} \right)
\label{eq:Gamma_def}, 
\end{equation}
where $B_\theta$ and $B_\phi$ are the local transverse components of the magnetic field in the local spherical coordinate basis ($\mathbf{e}_r,\,\mathbf{e}_\theta,\,\mathbf{e}_\phi$) with $\mathbf{e}_\theta$ pointing towards the South pole.
$\psi(r,\,\mathbf{n})$ is rotated 90$^\circ$ from the position angle of the plane-of-sky component of the local magnetic field.
\end{itemize}
This formulation assumes that the emissivity of the dust grains is constant through the Galaxy despite known evidences for variations of the dust spectral energy distribution (e.g., \citealt{Fin99}; \citealt{PlanckXI2014}; \citealt{Pel+21}).

As we do not consider multi-frequency analysis in this study, we are not affected by such variations which, in any case, should be small since we only produce maps corresponding to small 3D volumes to achieve high spatial resolution.

\begin{figure*}
    \centering
    \begin{tabular}{c c}
    \hspace{.2cm} filaments & \hspace{.4cm}sheets \\
    \includegraphics[trim={6.3cm 3.6cm 6.6cm 2cm},clip,width=.45\linewidth]{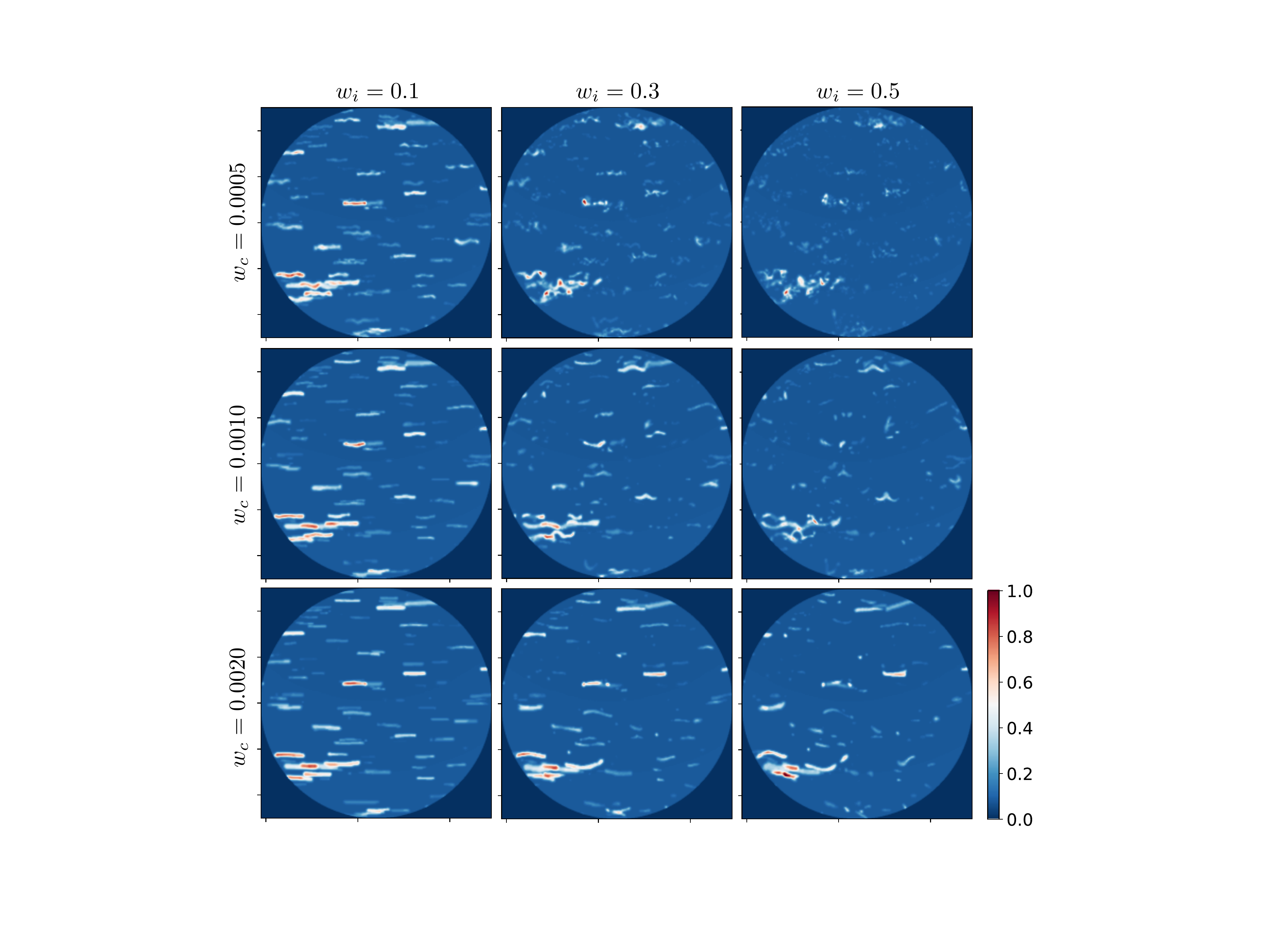} &
    \includegraphics[trim={6.3cm 3.6cm 6.6cm 2cm},clip,width=.45\linewidth]{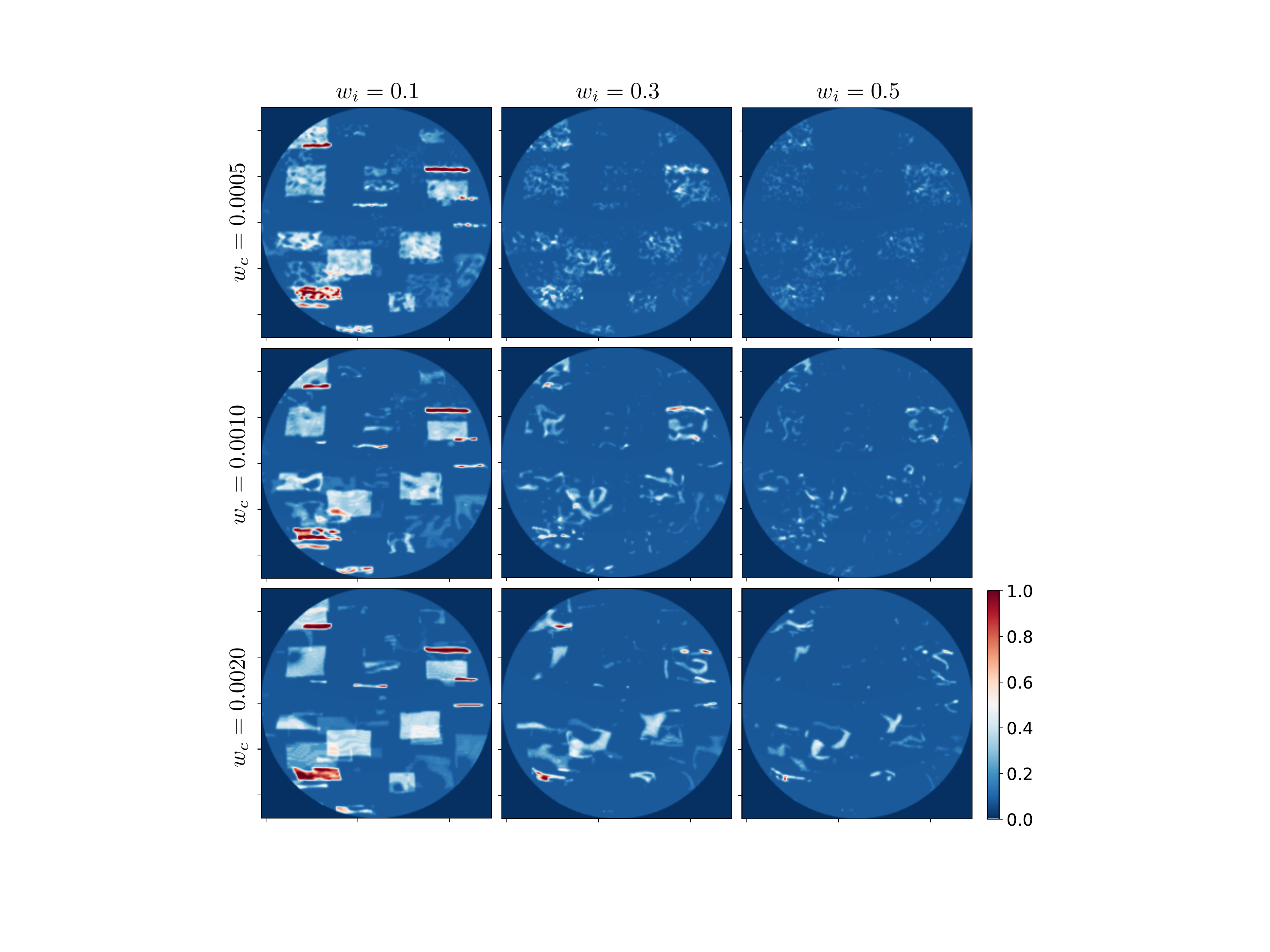}
    
    \end{tabular}
    \caption{ Column density ($K$)
    maps obtained for filament-like (left) and sheet-like (right) density structures with different wiggle parameters. The wiggle intensity ($w_i$) is constant in columns and increases from left to right and the wiggle correlation length ($w_c$) is constant in rows and increases from top to bottom.
    In our settings, $w_i$ and $w_c$ have units of [pc/(100 $\times$ 200)].
    The shape parameters of filaments and sheets are the same in each panel and read as follows (in pc). Filaments: $L \in [7,\, 25]$ and $R \in [1.5,\,4.5]$; Sheets: $L \in [5,\, 20]$, $R_{12} \in [10,\,20]$ and $R_{13} \in [1,\,3]$ (see Sect.~\ref{sec:dustclouds} and Table~\ref{tab:param_cloud} for parameter definitions and labels.)
    The circular regions displayed on those maps have an angular radius of 12.53$^\circ$.
    All maps share the same color scale, which is expressed in $10^{21}\,\mathrm{cm}^{-2}$.}
    \label{fig:fsh_wiggleparams}
\end{figure*}
\subsubsection{Populating the Galactic space with dust}

Within \textsc{Asterion} the Galaxy is populated by dust grains according to a smooth large-scale density distribution on top of which dust-cloud structures are added. The large-scale density distribution follows a parametric model that implements either the exponential disk model (ED) or the four logarithmic-spiral arms model (ARM4) discussed and adjusted to a full-sky Planck map in \cite{PMR21}. Their ARM4 model with best-fit parameter values is implemented as default in \textsc{Asterion} and we use that model in this work. The large-scale model fills the low-resolution grid which is then integrated to generate a dust column density map as seen from the Sun's location. A latitude profile is subsequently built and a normalization factor is computed such that it fits the well-known relation (\citealt{Kul1987}; \citealt{Hei03}):
\begin{equation}
    N_{\mathrm{H}_{\mathrm{I}}}(b) = \frac{3.7}{\sin(|b|)} \times 10^{20} \; [\textrm{cm}^{-2}],
\end{equation}
where $N_{\mathrm{H}_{\mathrm{I}}}$ is the column density of neutral hydrogen atoms and $b$ is Galactic latitude.

 The normalization factor fixes the proportionality in Eq.~\ref{eq:Columndensity1} leading the large-scale density distribution model to determine the number density of hydrogen atoms in any place in the Galaxy. The propagation of this normalization factor to Eqs.~\ref{eq:intensity} to~\ref{eq:DUSTEMISSION_U} implies that our polarization maps ($I$, $Q$, and $U$) are given in units of dust grain emissivity, as stated above.

The particle density assigned to the high-resolution box
(i.e., the OB)
is calculated from this large-scale model which is evaluated at the location of the center of the OB. Inside the OB, the total number of particles is determined by assuming that the matter in the OB is constant and therefore, scales with the OB volume. The total number of particles is then divided into two phases of the ISM which are the Warm Neutral Medium (WNM) and the Cold Neutral Medium (CNM) (e.g., \citealt{wolfire}).
In \textsc{Asterion} we assume that the WNM corresponds to a very diffuse component that follows the large-scale density distribution model, whereas the CNM forms the ISM structures (the clouds) which will appear brighter on the maps. The modeling of the clouds which we consider as being either filament-like or sheet-like structures is discussed in Sect.~\ref{sec:dustclouds}. The relative contribution of CNM and WNM to build the column density is a free parameter in \textsc{Asterion}. 
We currently do not consider the possibility for a contribution from  an unstable neutral medium phase (\citealt{Gho2017}; \citealt{Deb2020}).

Assuming a given number density of dust particles in clouds and given the volume of the dust clouds that we model, the number of clouds required to account for the mass of the CNM in the OB is automatically determined. The clouds are then placed randomly within the OB according to a uniform distribution. Both the particle number density and the morphology of the clouds are parameters that the user can tune. 

The orientation of clouds in the 3D space are not random. It is now well established observationally that the main axes of the projected shapes of ISM structures show orientations that are not independent of the orientation of the ambient plane-of-sky component of the magnetic field. Both orientations appear preferentially parallel or perpendicular, depending on their column density (e.g., \citealt{planck2016XXXII}; \citealt{Planck2016XXXV}; \citealt{clark_review}; \citealt{soler19}) and this alignment and misalignment may lead to differences in characteristics of the polarization power spectra (\citealt{planck2016XXXVIII}; \citealt{huffenberger}; \citealt{Cla+21}).
To control this correlation, \textsc{Asterion} allows the user to vary the angle  of each cloud in 3D with the local magnetic field. We refer to this angle as the offset angle which we denote $\omega$.
The implementation of magnetic field is discussed in Sect.~\ref{sec:Bfield}.
The correlation between dust density structures and local magnetic field is further discussed in Sect.~\ref{sec:turbulence}.

\begin{figure*}
    \centering
    \begin{tabular}{c c}
    \hspace{-.3cm} filaments & \hspace{-.3cm}sheets \\
    \includegraphics[trim={6.3cm 3.6cm 6.6cm 2cm},clip,width=.45\linewidth]{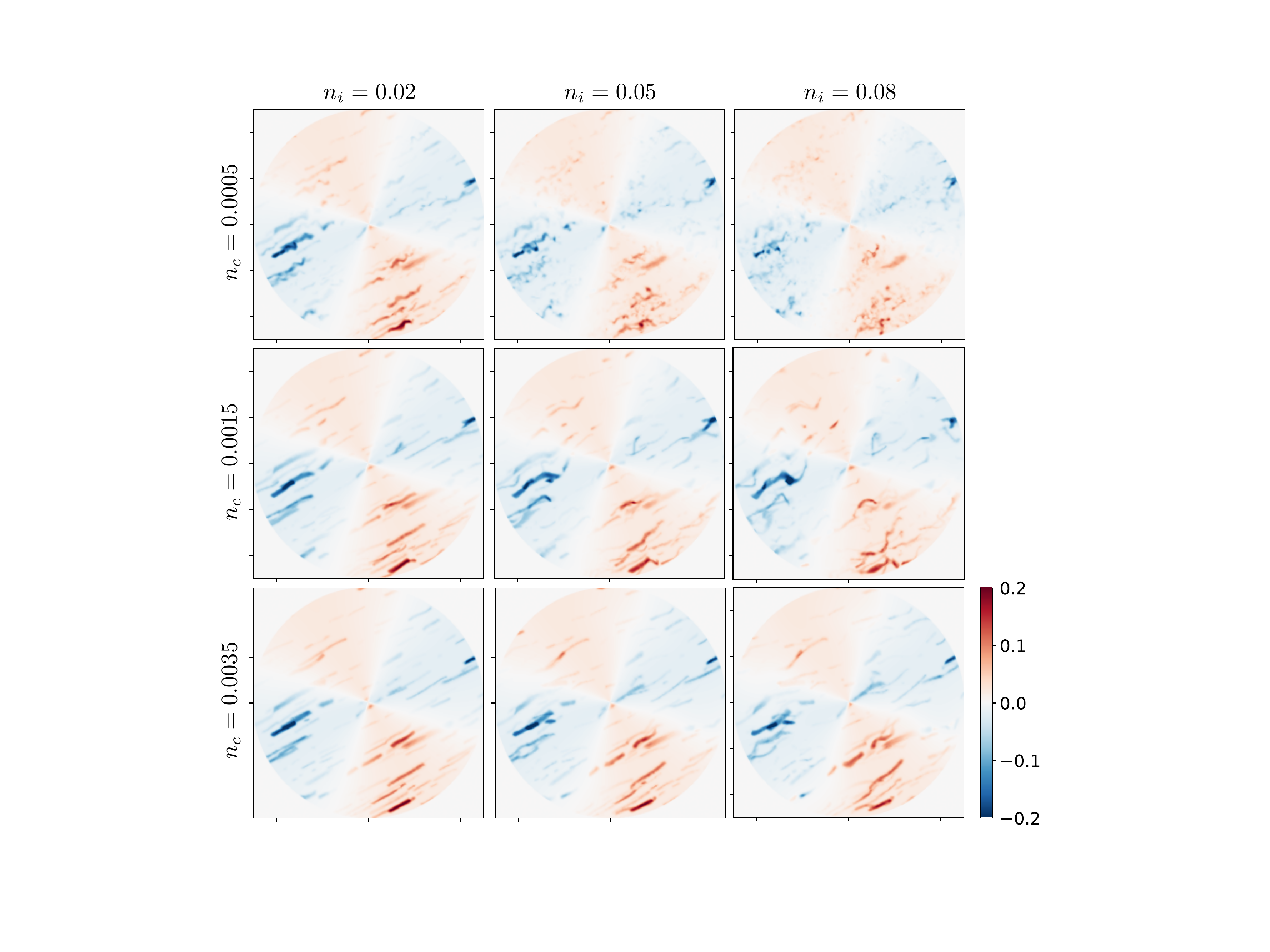} &
    \includegraphics[trim={6.3cm 3.6cm 6.6cm 2cm},clip,width=.45\linewidth]{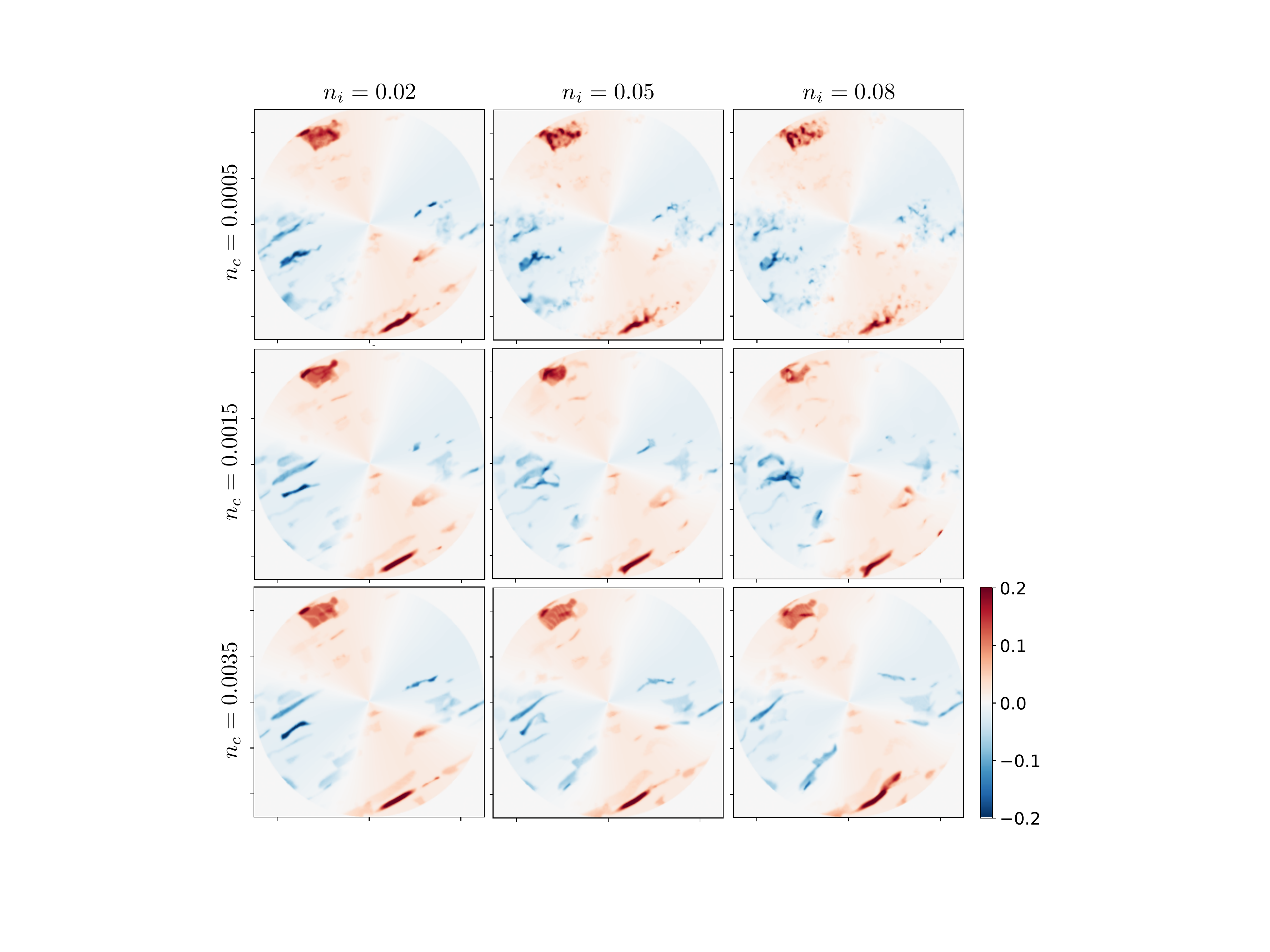}
    \end{tabular}
    \caption{
    Stokes $Q$ polarization maps obtained for filaments (left) and sheets (right) with different parameter values for the stochastic component of the magnetic field $B_{\rm{sto}}$. The component amplitude ($n_i$) is constant in column and increases from left to right and the correlation length ($n_c$) is constant in rows and increases from top to bottom.
    The circular regions, here centered on the North Galactic pole, have an angular radius of 12.53$^\circ$.
    Shape parameters, including those controlling the wiggles, are the same as in Fig.~\ref{fig:fsh_wiggleparams_v42} and are kept fixed between panels.
    All maps share the same color scale.}
    
    \label{fig:filshe_Bturbparams}
\end{figure*}

\subsubsection{Dust cloud structures}
\label{sec:dustclouds}
\textsc{Asterion} allows for the creation and visualization of dust clouds of the ISM that are either filament-like or sheet-like. 
The basic regular geometric shapes, which are then perturbed, are cylinders and flat rectangular parallelepipeds for filaments and sheets, respectively. The user can control the shape parameters (size, axis ratio, etc) and the density of dust grains within the structure. Generally, input parameters control the ranges of possible values and, internally, the software seeks random numbers from a uniform distribution to create any given implementation. This feature ensures intrinsic scatter which account for the fact that all dust clouds in the ISM are not identical.

These basic shapes are then perturbed in order to make their appearance more realistic, producing `wiggles'. For this purpose, \textsc{Asterion} relies on the use of Perlin noise. 
Perlin noise is a type of gradient noise used to increase appearance of realism in computer graphics simulating textures (\citealt{Per1985}). This technology is extensively used in video games and by the imaging industry. Two parameters control the properties of the noise, an amplitude ($w_i$) and a correlation length ($w_c$), that the user can change.

Finally, the density distribution within the OB is smoothed in 3D through a distance-weighted averaging of the densities from direct neighboring voxels. This smoothing is such that the total mass is conserved. It is applied to avoid sharp transitions between structures and their surroundings.

\smallskip

When the emission is integrated along the lines of sight (see Eq.~\ref{eq:intensity}), dust clouds can appear as illustrated in Fig.~\ref{fig:fsh_wiggleparams} where we show the intensity maps (Eq.~\ref{eq:intensity}) of a population of filaments (left panel) and sheets (right panel) obtained for different combinations of the wiggle parameters.
In these sky maps, the major axes of all the structures are horizontal. In the case of sheet-like clouds, the orientation of the minor axes are not bounded, which makes it possible to observe a cloud either edge-on or face-on. Among maps of filaments (sheets), only the wiggle parameters change, the 3D positioning of the clouds and their main orientation remain unchanged.
In Fig.~\ref{fig:fsh_wiggleparams}, the wiggle amplitude is constant through columns and increases from left to right while the wiggle correlation length is constant through rows and increases from top to bottom. The larger $w_i$ and the smaller $w_c$, the noisier the structures.

\subsubsection{Magnetic field}
\label{sec:Bfield}
A parametric model for the regular large-scale Galactic magnetic field is assumed in \textsc{Asterion}. 
Two models of the magnetic field are implemented: the Logarithmic Spiral Arm model of \cite{Page07} with the best-fit parameters obtained by \cite{PMR21} for a fit to dust polarization maps, and a four logarithmic-spiral arms model with plane-parallel field lines but with an azimuthal modulation of the strength that follow the large-scale dust density. 
The latter is the model that we use in this work despite the fact that it has never been tested against data. Our choice is only motivated by the fact that this model allows for an easier control of the relative orientation of the field lines with respect to the lines of sight of observation (see Sect.~\ref{sec:setup}).

\subsubsection{Mimicking MHD turbulence}
\label{sec:turbulence}

Following the description above, our simplistic approach to model the magnetized ISM does not couple matter density distribution and magnetic field as MHD equations do. To implement such a coupling in our toy-model and allow for the magnetic field lines to depart from the large-scale regular model, we proceed as follows.

First, we add a stochastic component to the magnetic field in the OB. The stochastic component is a random realization of a Perlin-noise vector field. This choice has the advantage of creating anisotropic fluctuations in the magnetic field directions. Two parameters control the stochastic component of the magnetic field; the amplitude of the fluctuations ($n_i$) measured in $\mu$G and a correlation length ($n_c$).
Then, we add the density structures (clouds). They are first placed and oriented as before and then modified (perturbed). To account for coupling we rely on the flux-freezing approximation (expected to be very accurate at the low density clouds we consider here for the high Galactic latitudes of interest), which dictates that matter can move freely along magnetic field lines. Therefore, a shift along the local (total) magnetic field line is imposed to each voxel of each cloud. The amplitude of the shifts and their spatial correlation is obtained from a scalar Perlin noise (the one that is used to generate the wiggles in the previous section) which is again controlled by the wiggle intensity ($w_i$) and wiggle correlation ($w_c$) parameters. Given our implementation, the structures can be squeezed or stretched along the magnetic field lines and compression and rarefaction can happen if the field lines converge or diverge. As a result, within individual structures, overdensities generally appear where magnetic field lines are more compressed as observed in the magnetized ISM (e.g., \citealt{Hey1983}; \citealt{Hei03}).

The effect of the stochastic component in the magnetic field is demonstrated in Fig.~\ref{fig:filshe_Bturbparams} where we show maps of the Stokes $Q$ (Eq.~\ref{eq:DUSTEMISSION_Q}) of a population of filaments (left) and sheets (right) located in a circular region towards the North Galactic pole. The orientation of the regular magnetic field crosses the sky regions making an angle of 28$^\circ$ with the horizontal which corresponds to the value of the pitch angle of the spiral arms in our model. In those panels, only the parameters of the stochastic magnetic-field component vary;
$n_i$ is constant in columns and increases from left to right with values of $n_i$ = 0.02, 0.05 and 0.08, and $n_c$ is constant in rows and increases from top to bottom with values of $n_c$ = 0.005, 0.015, 0.035.
The parameters of the wiggles are fixed to $(w_i,\,w_c) = (0.2,\, 0.001)$.

\begin{figure*}
    \centering
    \begin{tabular}{c c}
    \hspace{-.1cm} filaments & \hspace{-.1cm} sheets \\
    \includegraphics[trim={6.4cm 3.6cm 6.6cm 2cm},clip,width=.45\linewidth]{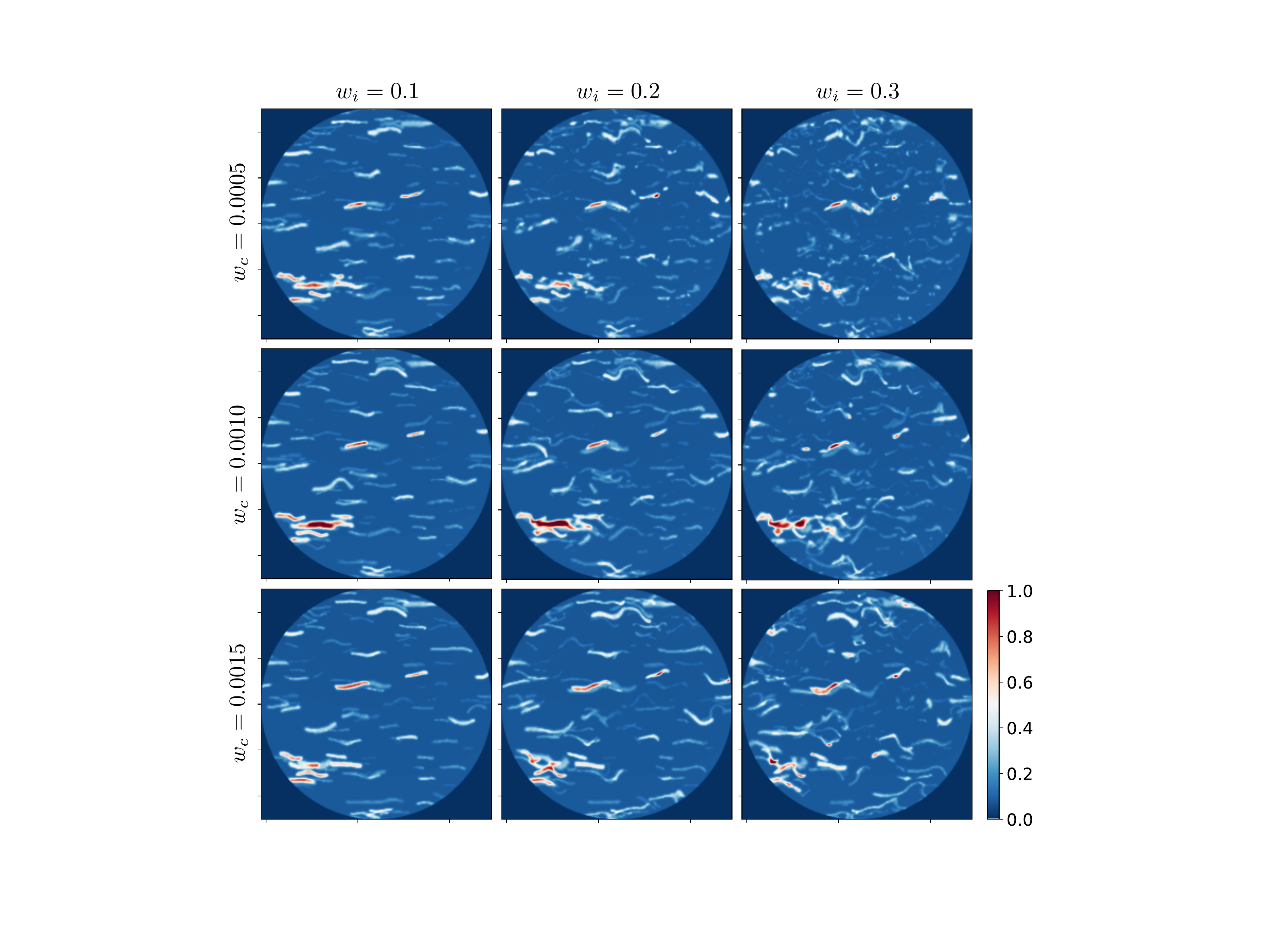} &
    \includegraphics[trim={6.4cm 3.6cm 6.6cm 2cm},clip,width=.45\linewidth]{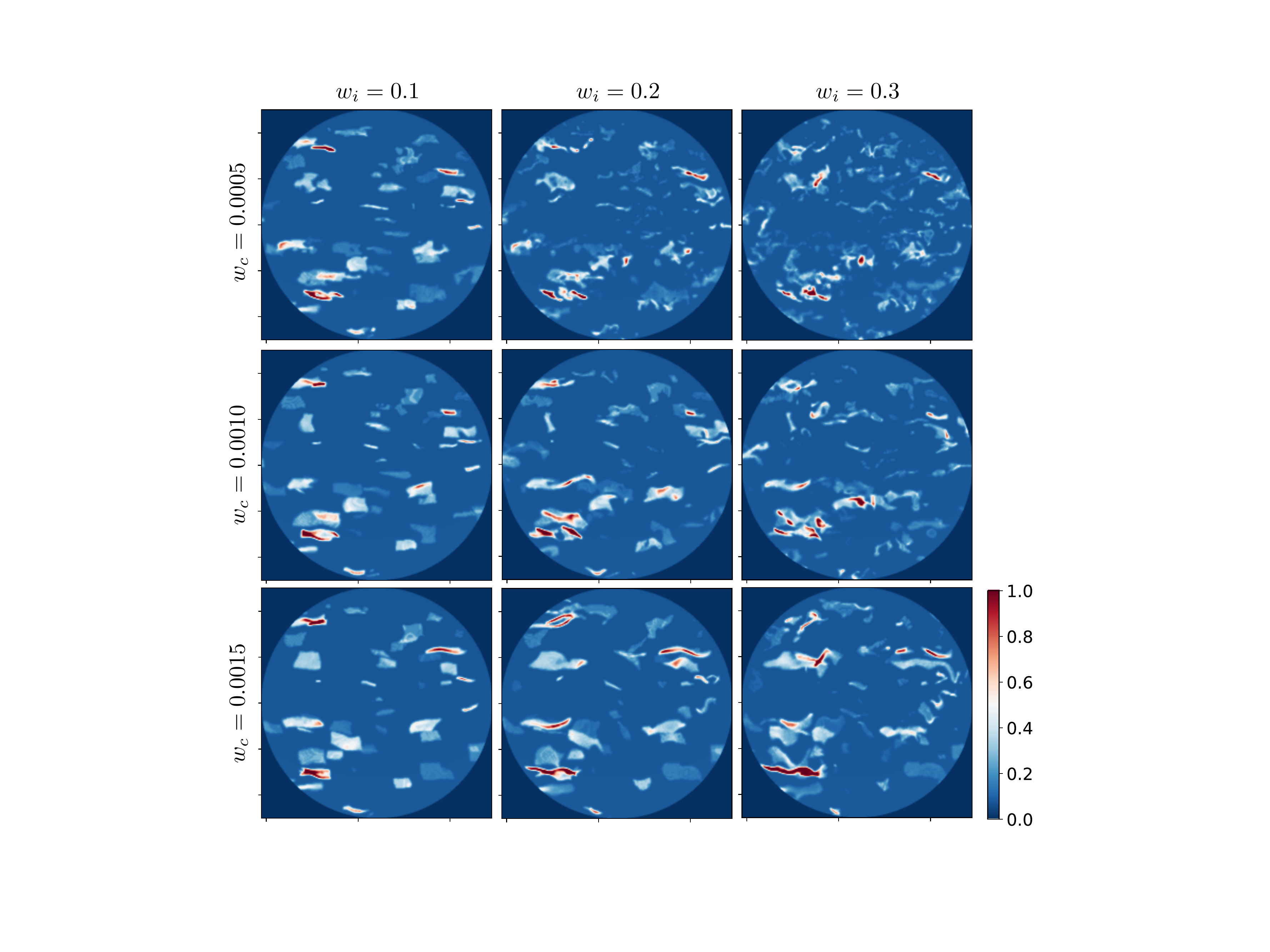}
    \end{tabular}
    \caption{
    Same as for Fig.~\ref{fig:fsh_wiggleparams} but for the case of non-regular magnetic field and density perturbation coupled to the local magnetic field 3D orientation. The realization of the magnetic-field stochastic component is the same for each panel and corresponds to $(n_i,\,n_c) = (0.05,\, 0.0015)$.}
    \label{fig:fsh_wiggleparams_v42}
\end{figure*}

The effects of varying the wiggle parameters to perturb the density structures in this coupling scheme are illustrated in Fig.~\ref{fig:fsh_wiggleparams_v42}. To produce those synthetic maps we proceed as for Fig.~\ref{fig:fsh_wiggleparams} (with $(\omega,\,\alpha) = (0^\circ,\,90^\circ)$) and fix the realization of the stochastic component in the magnetic field (including random seed value) to be the same with parameters $(n_i,\,n_c) = (0.05,\,0.0015)$.
The amplitude of the perturbation ($w_i$) is constant in a column and increases from left to right with values of $n_i$ = 0.1, 0.2 and 0.3 and the correlation length ($n_c$) is constant in a row and increases from top to bottom with values of $n_c$ = 0.0005, 0.0010, 0.0015.
Comparison of Figs.~\ref{fig:fsh_wiggleparams} and~\ref{fig:fsh_wiggleparams_v42} reveals, as expected, that the coupling of structure perturbations to magnetic field lead to significant difference in structures' morphology as the wiggles are not isotropic anymore.
As a side effect of our implementation of the coupling between magnetic field and matter density, structures may generally appear more aligned with the magnetic field than as imposed by the offset angle $\omega$, and may appear more elongated than when turbulence is switched off.
Both effects are in qualitative agreement with what MHD theory predicts (e.g., \citealt{Bra2013}).

\subsection{Synthetic data: Input parameters}
\label{sec:setup}

In this work, we want to study how the shape of ISM clouds and their alignment with the ambient magnetic field impact the measured polarization power spectra as well as to gauge the importance of the viewing angle.
This goal, together with the requirement to observe sky patches covering at least one per cent of the full sky (see next section), leads us to favour certain observer-OB configurations within \textsc{Asterion}.

First, we want to study a portion of the ISM typical of the intermediate-to-high Galactic latitude sky.
Therefore, we select a region placed at $(l_\odot,\,b_\odot) \approx (180^\circ,\,44^\circ)$ so that the entire observed region, as seen from the Sun, is above latitude 30$^\circ$. We want to resolve density structures at the pc scale and therefore decide for an OB size of 200 pc (the voxels are 0.78 pc on a side). To achieve an observed sky area of 1\% of the full sky, the circle defined by the intersection of the observation cone and the back side of the OB has to have an angular size of $\approx$ 25$^\circ$ and, therefore, the center of the OB has to be placed at 350 pc from the observer.
The distance between the Sun and the OB is $d_{\odot} \approx$ 370 pc and the observer will not be at Sun position.

Second, the fact that the observables returned by \textsc{Asterion} reflect only the portion of the ISM that is encompassed by the OB makes it possible to study the same ISM configuration from different viewpoints.
We take advantage of this feature to control values of the inclination angle between the (regular) magnetic field orientation at the center of the OB and the line of sight toward the center of the OB (the angle $\alpha$ in Eqs.~\ref{eq:intensity} to~\ref{eq:DUSTEMISSION_U}).
We notice that in general $\alpha$ may vary across the outputs.
The large-scale regular magnetic field model implements magnetic field lines following a logarithmic-spiral pattern with a constant pitch angle of 28$^\circ$ and with zero out-of-plane component.
Therefore, we choose to place the observer at a height $z=$ 257.1 pc ($= d_\odot \,\sin(b_\odot)$) above the Galactic disk. We move the observer in this plane while keeping the same distance to the OB center in order to explore the effect of the inclination angle on the observations. In this case, all observations made by the observer are at latitude 0$^\circ$ and moving the observer around the OB changes the longitude only. We sample the inclination angles from 90$^\circ$ (magnetic field in the plane of the sky) to 0$^\circ$ (magnetic field along the line of sight pointing to the center of the OB) with step of 15$^\circ$. The angle that the main axes of the clouds make with the orientation of the magnetic field is the other parameter that we want to explore. We vary this angle from 0 to 90$^\circ$ in 15$^\circ$ step.
To allow us to focus only on the effects from the shape, the viewing angle and the misalignment between structures and magnetic field, we start our study focusing on simulating data without the addition of the  stochastic component in the magnetic field and the correlations between density perturbations and magnetic field. Another set of synthetic maps is generated in Sect.~\ref{sec:withturbulence} where we study the effects from those variables.

\smallskip

To decide on a default setting for the large number of free parameters allowed by \textsc{Asterion}, we rely on observational constraints and on visual inspection of the resulting column density maps so that the latter appear as realistic as possible.

First, according to \cite{Hei03}, we consider that the CNM (the cloud structures) must account for 39\% of the gas mass and that the remaining 61\% is attributed to the WNM (the large-scale density distribution).
Second, based on \cite{spitzer}, we consider that the particle number density in clouds must be chosen randomly for each cloud in the range of 10 to 70 $\mathrm{cm}^{-3}$.

Following \cite{Heiles1976}, the typical length of clouds is expected to be around 20 pc with a width in an approximate ratio of 10:1.
We explored \textsc{Asterion}'s parameter space around those values also varying the parameters describing the wiggles (the fluctuations in the density distribution within the clouds), and the values or range of values of all parameters we choose are listed in Table~\ref{tab:param_cloud}. The wiggle parameters are fixed to the same values for filaments and sheets. 
For filaments, the length is drawn from a uniform random distribution in the range 7 to 25 pc while the thickness varies uniformly from 1.5 to 4.5 pc. For sheets, the length is drawn from a uniform random distribution in the range from 5 to 20 pc, the width of the sheets takes random values in the range from 10 to 20, and the thickness (the height) in the range 1 to 3.

\smallskip

We adopt the observer-OB configurations presented above together with the chosen settings for the cloud morphology and generate 10 realizations of the magnetized ISM for each view point by changing the random seed in \textsc{Asterion}.
To summarize, for both filaments and sheets, we explore seven values for the inclination angle, seven values for the offset angle (between the structures' main axis and the magnetic field) and we generate 10 realizations for each couple of angles in order to infer the variance induced by the positioning of the clouds and the scatter from the specific choices of cloud size and shape and small-scale fluctuations. Therefore, we generate a set of 490 observations for each. An example of $I$, $Q$ and $U$ polarization maps is shown in Fig.~\ref{fig:I_map_ambiguity} for clouds being filaments (top) or sheets (bottom) with major axis aligned with the ambient magnetic field lines; the latter making an angle of $90^\circ$ with the line of sight for the filament output and $60^\circ$ for the sheet output. Since the magnetic field runs horizontally, most of the polarization signal is observed in positive Stokes $Q$.

However, given the large area of our output maps, some signal is also observed in Stokes $U$ far from the center of the maps because of projection effects. Even for the case of a uniform magnetic field in the OB, the effective inclination and position angles of the magnetic field with respect to the line of sight and in the plane of the sky vary in the image. The smaller the inclination angle, the larger the variation in the position angle.

\begin{table}
\centering
\caption{Parameters and values kept constant in our toy-models for filaments and sheets.}
\label{tab:param_cloud}
{\small{
\begin{tabular}{llll}
\hline
\hline \\[-1.5ex]
shape & parameters & label & value (range)			\\[.5ex]
\hline
\\[-1.ex]
Filaments 	
		& Length [pc]       & $L$   & $[7.0-25.0]$	\\
		& Thickness [pc]    & $R$   & $[1.5-4.5]$ \\

\\[-1.5ex]
Sheets
		& Length [pc]       & $L$       & $[5.0-20.0]$	\\
		& Width [pc]        & $R_{12}$  & $[10.0-20.0]$ \\
		& Thickness [pc]    & $R_{13}$  & $[1.0-3.0]$   \\
	
\\[-1.5ex]
Both
		& Background Density    &   & 	$0.61$	\\
		& Particle density [cm$^{-3}$] & $n_{\rm{d}}$ & $[10.0-70.0]$ \\
		& Wiggle Intensity      & $w_i$ & $0.003$	\\
		& Wiggle Correlation    & $w_c$ & $0.001$ \\
\\[-1.5ex]
\hline
\end{tabular}
}}
\label{tab:values}
\end{table}

\subsection{Polarization power spectra}

\subsubsection{Formalism and main characteristics}
The analyses of polarization power spectra are generally carried out using pseudo angular power spectra defined as:
$\mathcal{D}_\ell^{XY} = \ell (\ell +1) \, C_\ell^{XY} / (2\pi)$ where $C_\ell^{XY}$ are the angular (auto- or cross-) power spectra and where $X$ and $Y$ refer either to $T$, $E$ or $B$ (e.g., \citealt{bracco}).

The $E/B$ power asymmetry is measured through the $\mathcal{R}_{EB}$ ratio, which is obtained by averaging the ratio of the auto-power spectra $\mathcal{D}_\ell^{EE}$ and $\mathcal{D}_\ell^{BB}$ over a specified multipole range
\begin{eqnarray}
    \mathcal{R}_{EB} \equiv \left< \frac{\mathcal{D}_\ell^{EE}}{\mathcal{D}_\ell^{BB}} \right>,
\end{eqnarray}
where $\left\langle \cdot \right\rangle$ stands for the mean over multipole bins.

To quantify the correlation between the $T$ and $E$ power spectra we use the normalized parameter $r^{TE}_\ell$ introduced by \cite{Cal2017}. $r^{TE}_\ell$ takes values $1$, $-1$ and $0$ in case of perfect correlation, perfect anti-correlation and absence of correlation, respectively, and is defined as
\begin{equation}
\label{eq:r_XY}
r^{TE} = \left\langle r^{TE}_\ell \right\rangle = \left\langle \frac{C_\ell^{TE}}{\sqrt{C_\ell^{TT} \, C_\ell^{EE}}} \right\rangle \; .
\end{equation}
The correlation between the power spectra $T$ and $B$ discussed in Sect.~\ref{sec:discussion} is also computed through the correlation coefficient $r^{TB}$ obtained by substituting $E$ for $B$ in Eq.~\ref{eq:r_XY}.
Additionally, visual inspection of the power spectra informs us that most of the $TT$, $EE$ and $BB$ auto-power spectra present an apparent power-law dependence in $\ell$ in the range $\ell \in [100,\, 500]$. Consequently, we decide to adjust a power-law power spectrum model of the form 

$D_\ell^{XX}=A^{XX}_{80}(\ell/80)^{a_{XX}+2}$, where $X \in \{E, B\}$,
and to characterize the spectra through the values of $A^{XX}_{80}$ and $a_{XX}$.

\subsubsection{Computation from synthetic maps}

\begin{figure*}[!ht]
    \centering
    \rotatebox{90}{\hspace{1.4cm} $Filaments$}
    \includegraphics[trim={0cm 0cm 0cm 0cm},clip,height=4.cm]{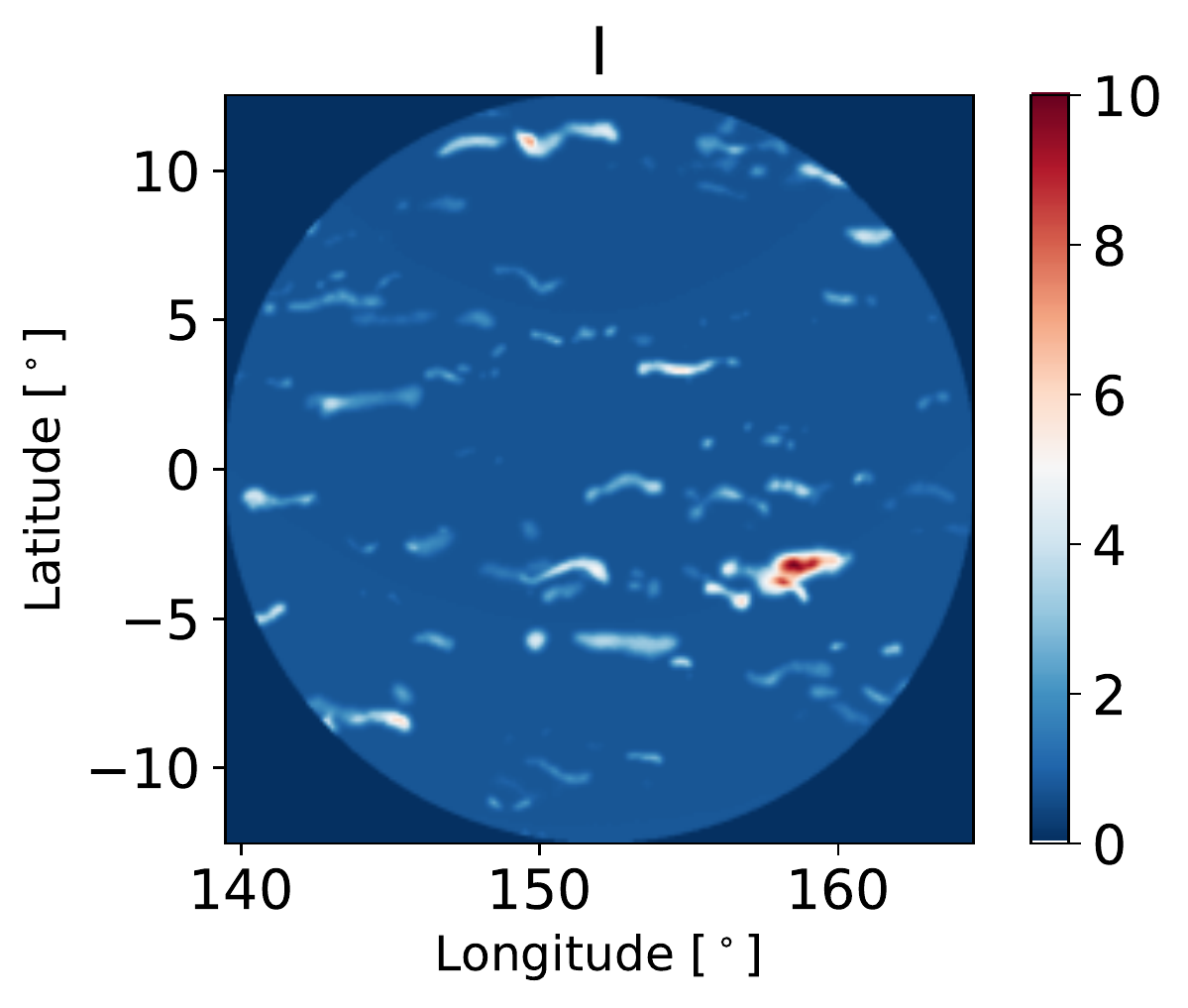}
    \includegraphics[trim={2.05cm 0cm 2.8cm 0cm},clip,height=4.cm]{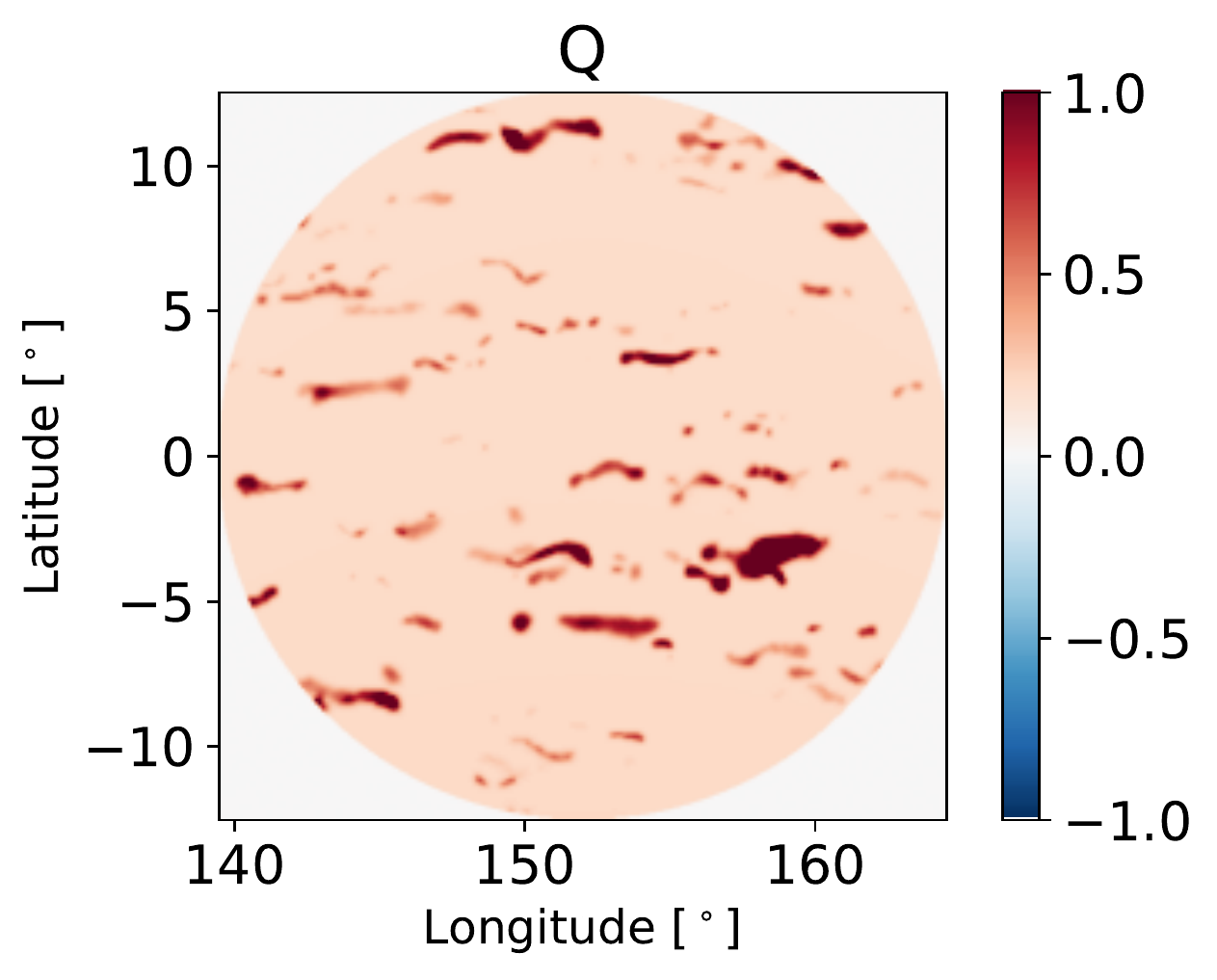}
    \includegraphics[trim={2.05cm 0cm 0cm 0cm},clip,height=4.cm]{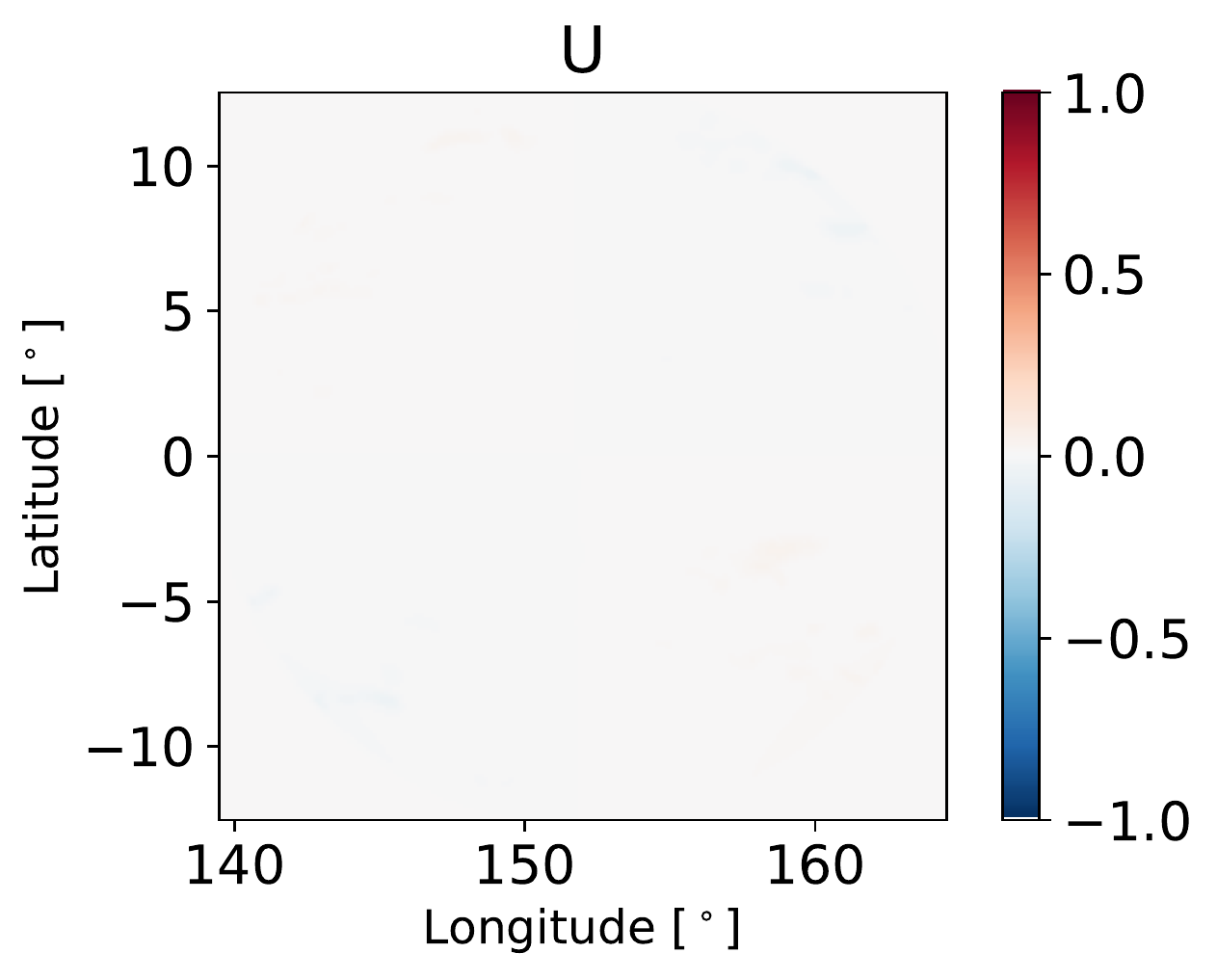}
    \includegraphics[trim={0cm 0cm 0cm 0cm},clip,height=3.7cm]{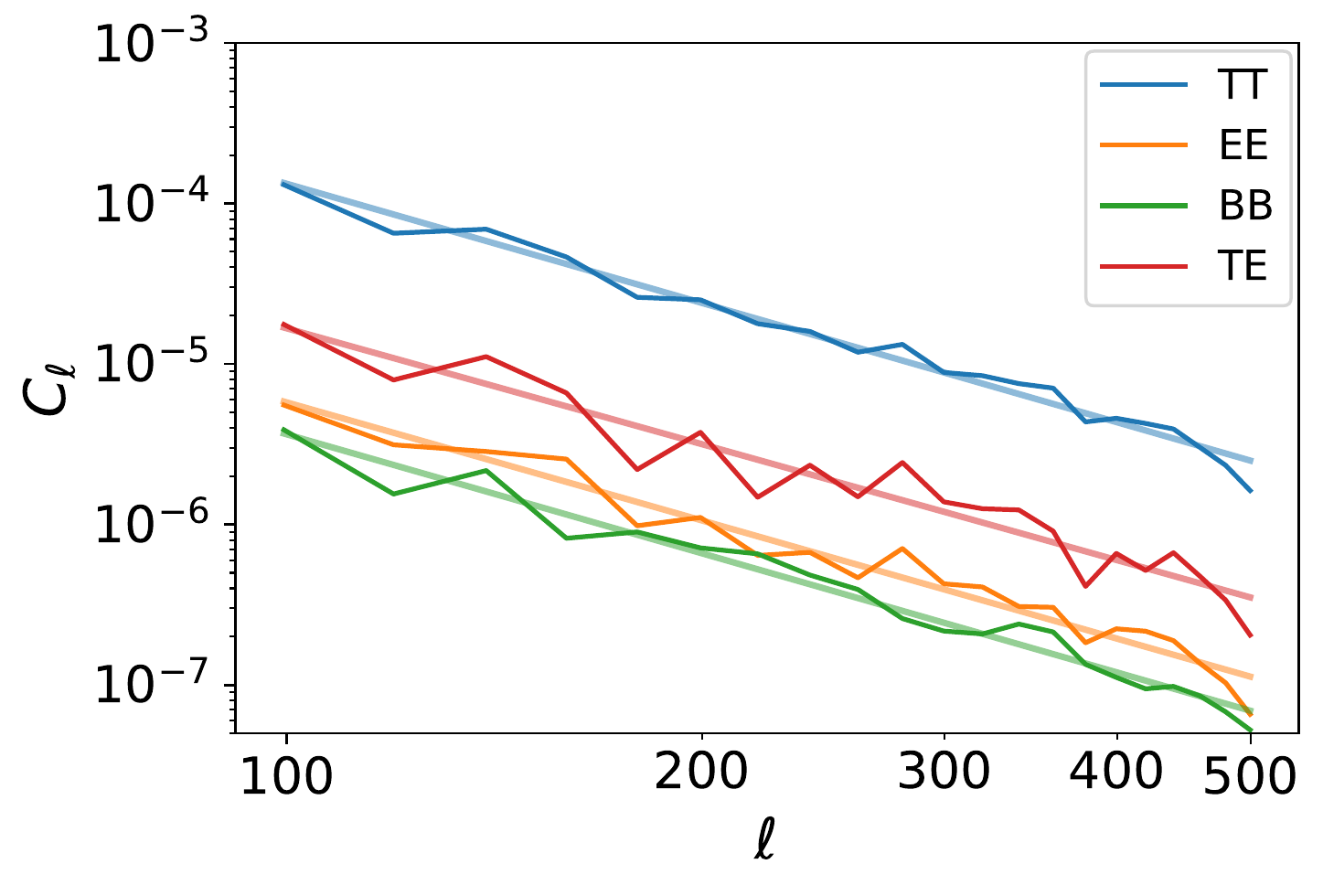}
    \\
    \rotatebox{90}{\hspace{1.4cm} $Sheets$}
    \includegraphics[trim={0cm 0cm 0cm 0cm},clip,height=4.cm]{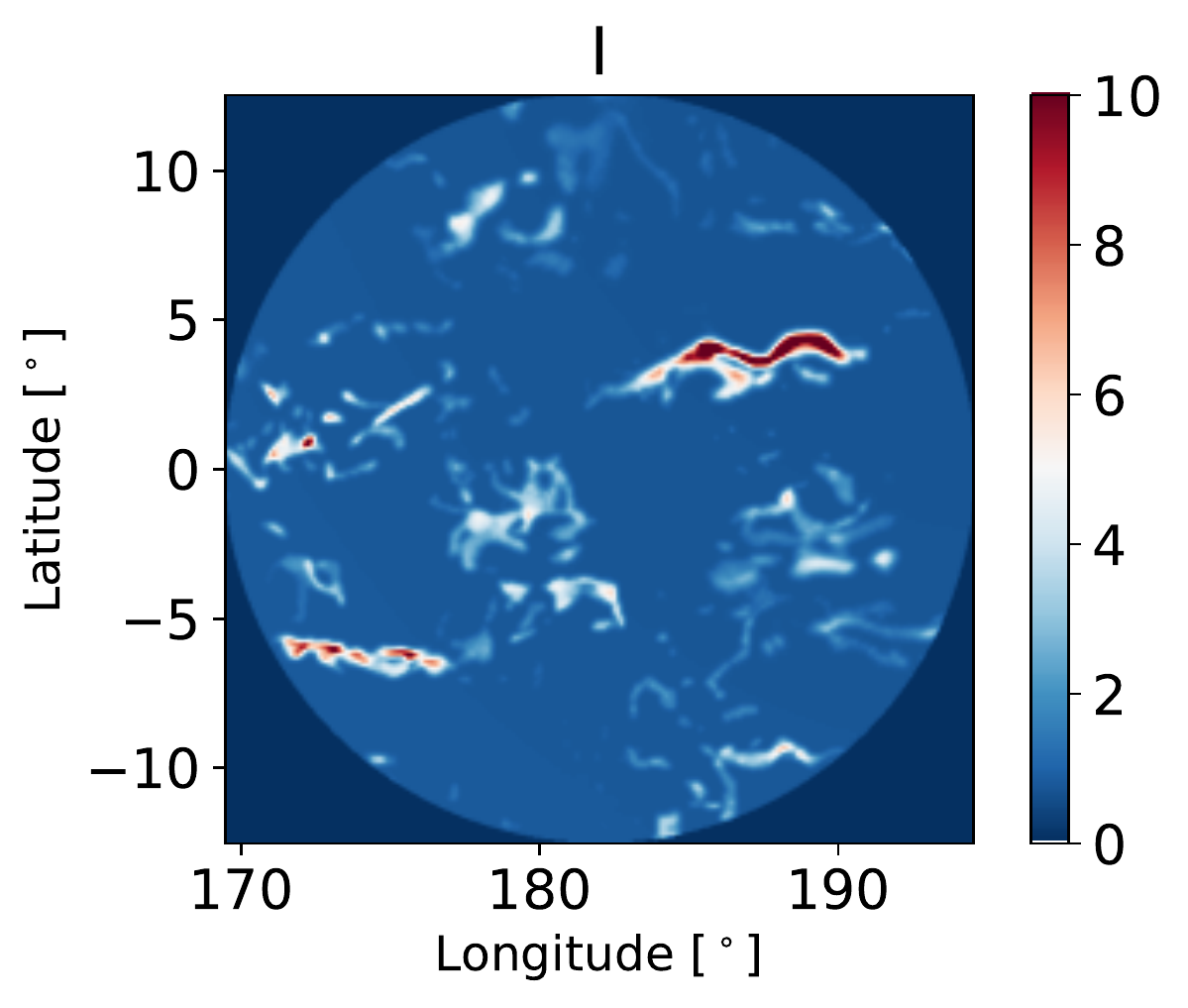}
    \includegraphics[trim={2.05cm 0cm 2.8cm 0cm},clip,height=4.cm]{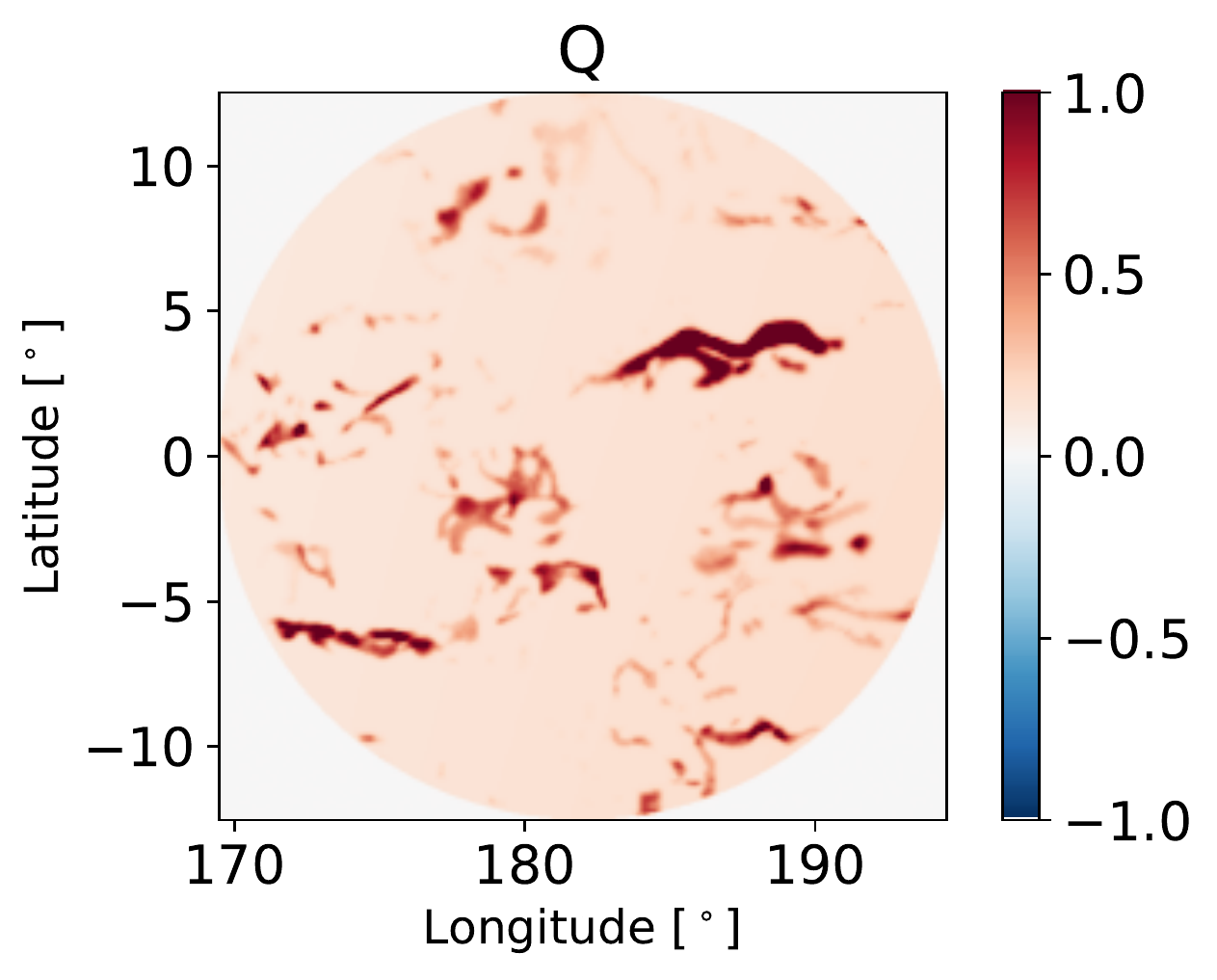}
    \includegraphics[trim={2.05cm 0cm 0cm 0cm},clip,height=4.cm]{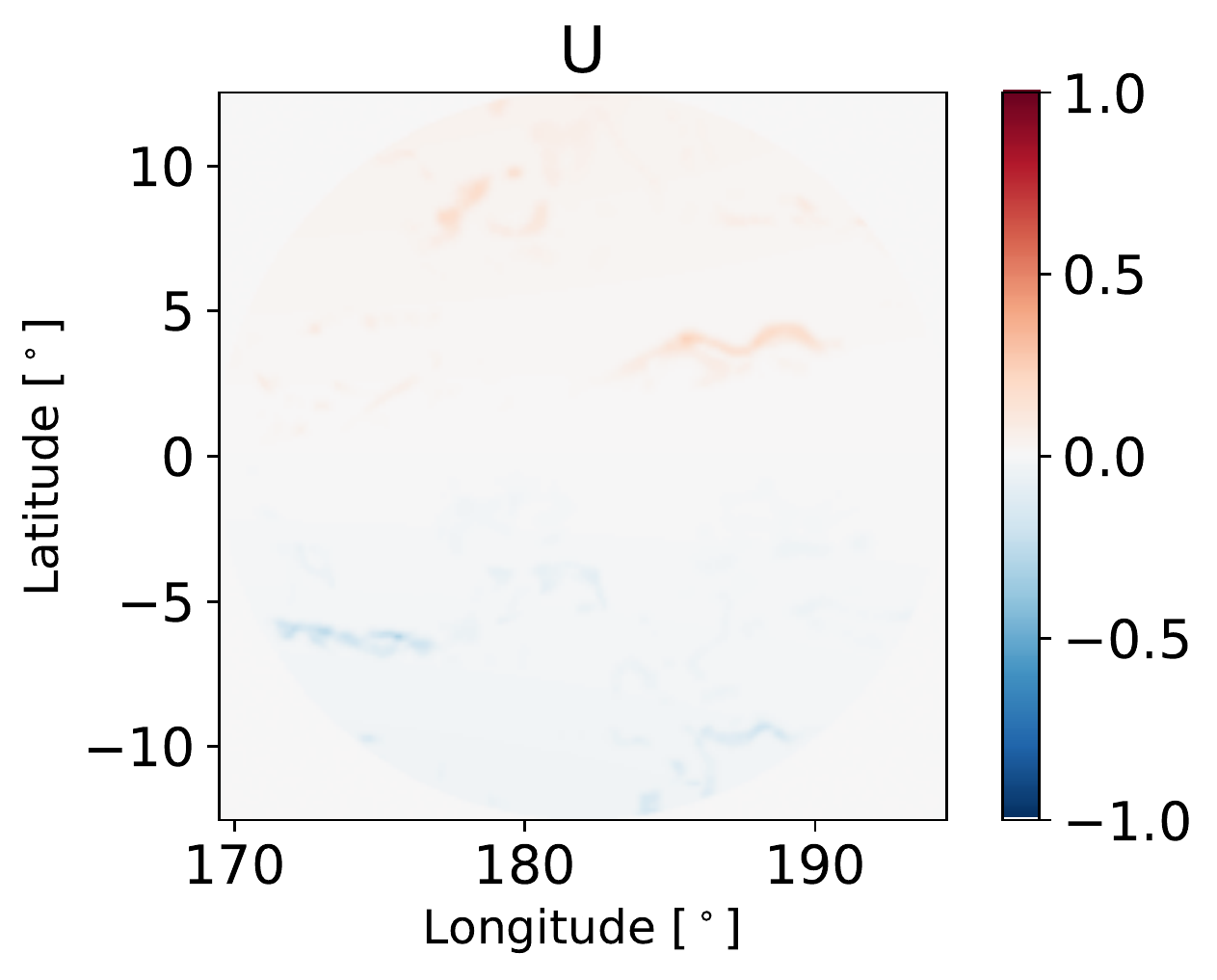}
    \includegraphics[trim={0cm 0cm 0cm 0cm},clip,height=3.7cm]{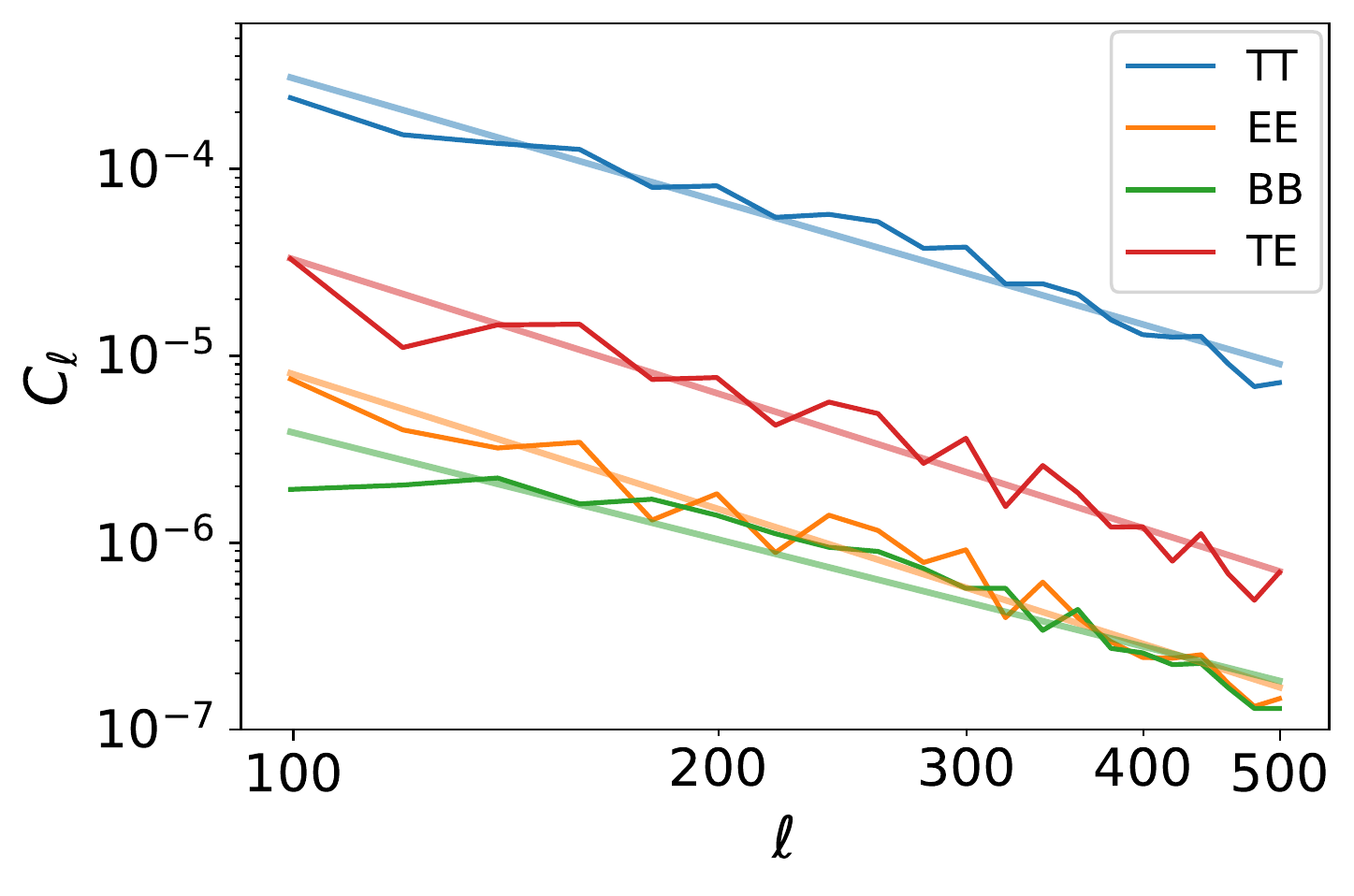}
    \caption{From left to right: Intensity ($I$), $Q$ and $U$ polarization maps and polarization power spectra obtained for filament-like (top) and sheet-like (bottom) clouds. The values of quantities extracted from the power spectra are $\mathcal{R}_{EB}=1.69$, $r^{TE}=0.65$, $a_{TT}=-2.47$, $a_{EE}=-2.45$, $a_{BB}=-2.47$ for filaments and $\mathcal{R}_{EB}=1.36$, $r^{TE}=0.61$, $a_{TT}=-2.19$, $a_{EE}=-2.40$, $a_{BB}=-1.91$ for sheets.}
    \label{fig:I_map_ambiguity}
\end{figure*}
We use the \texttt{Xpol} code\footnote{\url{https://gitlab.in2p3.fr/tristram/Xpol}} (\citealt{tristram}) to compute the polarization power spectra of the dust polarized sky and account for incomplete sky coverage.
The reliability of \texttt{Xpol} at estimating polarization power spectra has been tested for sky area larger or equal to one per cent of the full-sky\footnote{Tristram 2020, \textit{private communication}}. This limit fixes the minimum angular size of the output that we generate using \textsc{Asterion}. This choice, together with the spatial resolution that we want to achieve in the OB, further sets the size of the OB and its distance to the observer (see previous section).

In practice, we project \textsc{Asterion}'s outputs onto a full-sky HEALPix map with resolution parameter $N_{\rm{side}} = 2048$. Most of the map is empty. We then define a circular mask of radius 11.5$^\circ$ centered on the outputs' center. This choice allows us to retain only the sky area that is free of artifacts from ray-tracing the volume-limited cubic OB. The sightlines that are kept pass through both the front and back faces of the OB, as seen by the observer.
To avoid power leakage we further apodize (smooth) the mask using a 0.5 degree FWHM beam. This mask is applied to \textsc{Asterion}'s outputs when projected on HEALPix map. The effective sky fraction of the mask ($f_{\rm{sky}}$) is 1\% of the full-sky. Using Monte Carlo simulations we checked that our mask does not introduce bias or leakage in the power spectrum estimation following the same validation procedure as in (\citealt{PelNT22}).

For the computation of the power spectra there are some constraints imposed for the multipole moments. The maximum value of $\ell$ is defined by the effective angular resolution achieved in \textsc{Asterion}'s outputs. Visual inspection of several power spectra informs us that above $\ell = 500$ there is a sudden drop of power, reflecting the effects of the limited spatial resolution achieved in the OB and the smoothing of density structure.

The threshold at low multipole values comes from uncertainties in power spectrum estimation.

Since our synthetic maps do not contain observational noise, we consider only the sampling variance as a source of uncertainty in our power spectrum estimates. The latter is linked to the number of unmasked pixels and their spatial arrangement on the sky. The analytical estimate from the sampling variance is given by (\citealt{planck16XXX}):
\begin{equation}
   \sigma^{XX}_{C_\ell}= \sqrt{\frac{2}{(2 \ell+1)\,f_{\mathrm{sky}}\, \Delta\ell_{\rm{bin}}}}\,C^{XX}_{\ell} \; ,
\end{equation}
where $f_{\rm{sky}}$ and $\Delta \ell_{\rm{bin}}$ are the sky fraction and the width of the multipole bins used to estimate the power spectra, respectively.
We choose to reject power spectrum estimates for multipole $\ell$ below 100 so as to ensure that $C_\ell / \sigma_{C_\ell} \geq 3$ always.

To summarize, we use \texttt{Xpol} to estimate the polarization power spectra from \textsc{Asterion} outputs in the multipole range $\ell \in [100,\, 500]$ adopting bins of width 20. We compute the $TT$, $EE$, $BB$ and $TE$ spectra for every output of our sample and characterize them through their $\mathcal{R}_{EB}$ and $r^{TE}$ values. An example of such a set of polarization power spectra is shown in Fig.~\ref{fig:I_map_ambiguity} along with the corresponding polarization maps.


\section{Analysis}
\label{sec:analysis}
In this section, we explore the dependence of the values of $\mathcal{R}_{EB}$ and $r^{TE}$ with respect to (i) the offset angle between the clouds' long axes and the magnetic field ($\omega$) and (ii) the inclination angle of the magnetic field with the line of sight, for both filament-like and sheet-like clouds. We vary both angles from 0$^\circ$ to 90$^\circ$ with a step of 15$^\circ$. Since the exact values of $\mathcal{R}_{EB}$ and $r^{TE}$ depend sensitively on the specific choice of parameters used to produce the synthetic maps, we are primarily interested in the trends that they exhibit as a function with $\omega$ and $\alpha$. 

\subsection{$\mathcal{R}_{EB}$ and $r^{TE}$ dependence on $\omega$}
Here we infer the dependence of $\mathcal{R}_{EB}$ and $r^{TE}$ on the angle that the major axes of the structure makes in three dimensions with the local magnetic field orientation.

\begin{figure*}
    \centering
    $\omega$ dependence of $\mathcal{R}_{EB}$ and $r^{TE}$\\[-.3ex]
    \includegraphics[trim={2.2cm 4.1cm .4cm .9cm},clip,width=.8\textwidth]{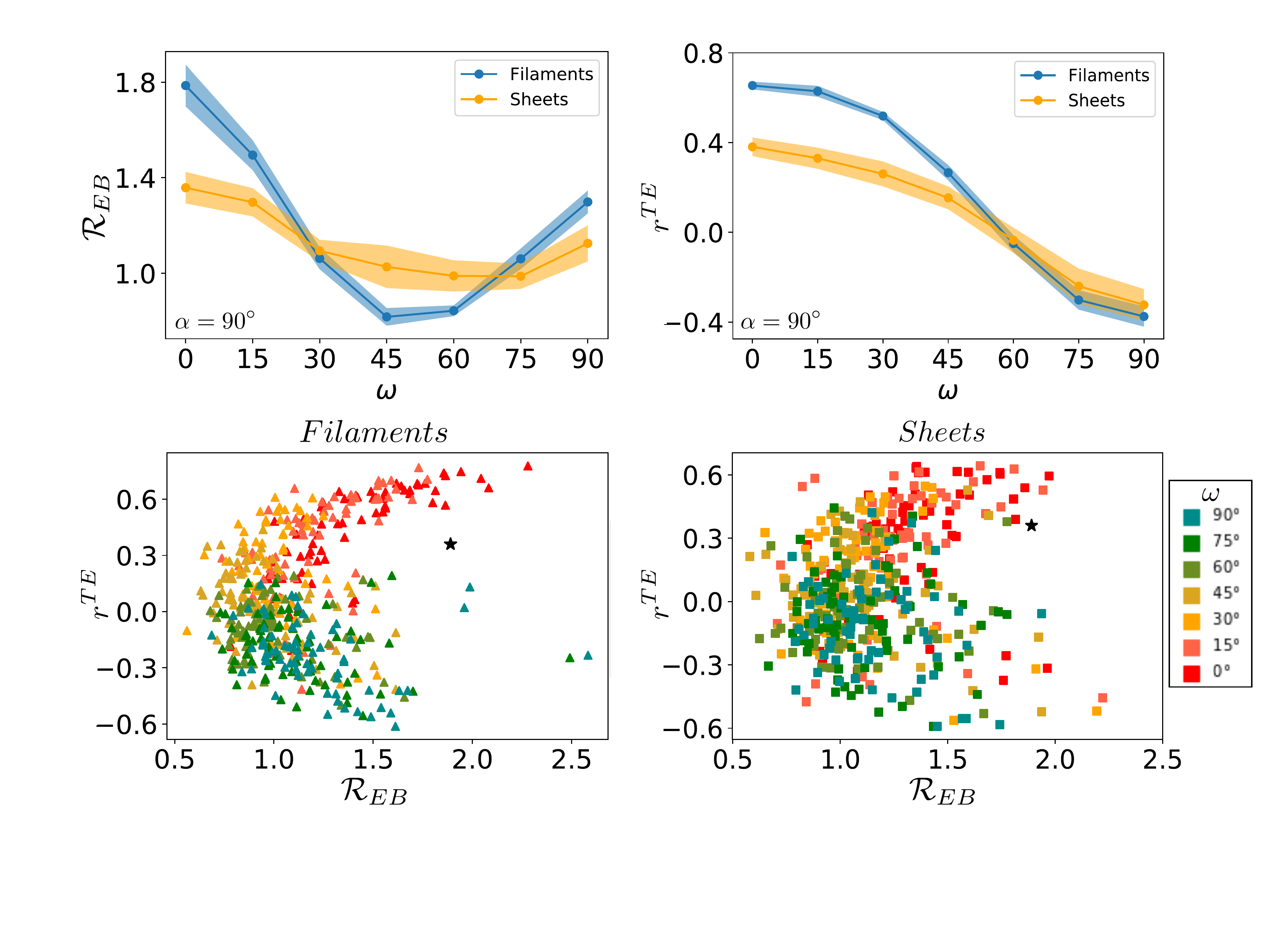}
    \caption{Dependence of $\mathcal{R}_{EB}$ and $r^{TE}$ values on $\omega$, the misalignment angle between magnetic field's orientation and structures' main axes for both filament and sheet structures' shapes. Top row shows $\mathcal{R}_{EB}$ (left) and $r^{TE}$ (right) versus $\omega$ for $\alpha = 90^\circ$ for both filaments (blue) and sheets (orange). The shaded areas include 68\% of the data around the median (thick lines) obtained through 10 random realizations of the same setup. Bottom row shows scatter plots of $(\mathcal{R}_{EB},\,r^{TE})$ pairs for filaments (left) and sheets (right) color-coded according to the $\omega$ value and obtained for the complete set of synthetic map. For visual reference, the black star in the bottom row indicate the fiducial Planck values $(\mathcal{R}_{EB}=1.89,\,r^{TE}=0.36)$.}
    \label{fig:REB-rTE_vs_omega-s-f}
\end{figure*}

We first select the synthetic maps created with the magnetic field in the plane of the sky ($\alpha=90^\circ$) and we sort them according to the offset angle that the major axes of the structures make with the local magnetic field. 
In the top row of Fig.~\ref{fig:REB-rTE_vs_omega-s-f}, we show the dependence of $\mathcal{R}_{EB}$ (left) and $r^{TE}$ (right) as a function of $\omega$, for both shape families of clouds.

For filament-like clouds, we observe that $\mathcal{R}_{EB}$ as a function of $\omega$ follows a parabolic trend starting from a maximum when clouds' major axes are perfectly aligned with the magnetic field ($\omega = 0^\circ$), then decreases to reach a minimum toward $\omega = 45^\circ$ and increases back for structures perpendicular to the magnetic field ($\omega = 90^\circ$). This trend reflects well the generally accepted picture according to which $E$ modes are maximized for structures perfectly parallel or perpendicular to the magnetic field whereas $B$ modes dominate when (projected) linear structures make an angle of $45^\circ$ with it (e.g. \citealt{zaldarriaga}; \citealt{huffenberger}; \citealt{Cla+21}). Obtaining this result, for filaments, validates our modeling and analysis pipeline.
Still for filament-like clouds, we observe that $r^{TE}$ as a function of $\omega$ follows a decreasing squared-cosine trend, even showing negative values for large offset angles ($\omega \geq 60^\circ$). This trend is reminiscent of the result obtained for individual straight cylinders (\citealt{huffenberger}) and is understood by the same argument as above. When $E$ modes are reduced due to misalignment, the corresponding $E$ map systematically looses its correlation with the $T$ map. The two even get anti-correlated when the structures are perpendicular to the field since $E$ modes become negative in such a configuration.

For sheet-like clouds, the same parabolic trend is recovered for $\mathcal{R}_{EB}$ as a function of $\omega$ but with a smaller absolute difference between the extrema than in the case of filament-like structures.
This is also true for the squared-cosine trend of $r^{TE}$ versus $\omega$.
We understand these observations by the fact that sheets have one additional degree of freedom in their orientation. The fact that they could be seen either face-on, edge-on or with any intermediate angles, even though the orientation of their principal axes is fixed as compared to the magnetic field orientation (here parallel to it), has the effect of reducing the imbalance between $E$ and $B$ modes. Even in the case the sheets' major axes are parallel to the magnetic field, when projected on the sky, wiggles of the sheet-like structures may show appreciable angles with the magnetic field lines, contributing to the production of $B$ modes rather than $E$ modes. This reduces both the $\mathcal{R}_{EB}$ and $r^{TE}$ maximum values.

For both sheet- and filament-like structures, we observe that the absolute difference between extrema in the trend of $\mathcal{R}_{EB}$ versus $\omega$ is reduced when the inclination angle decreases, for example when the field lines depart from the plane of the sky. This is again understood by projection effects and is further explored in Sect.~\ref{alphadep}.

\smallskip

In the bottom row of Fig.~\ref{fig:REB-rTE_vs_omega-s-f}, we show the scatter plots of $r^{TE}$ versus $\mathcal{R}_{EB}$ obtained for filament-like clouds (left) and sheet-like clouds (right), for all offset and inclination angles. The scatter plots are color-coded according to the offset angles.
These plots offer a different and generalized view of the trends observed in the top panels which were restricted to the case $\alpha=90^\circ$.

Particularly, for the case of filament-like clouds (left), we notice a tail in high $\mathcal{R}_{EB}$ and $r^{TE}$ values which corresponds to small offset angles ($\omega\lesssim 30^\circ$). Then, as $\omega$ increases up to $\approx 45^\circ$, both the values of $\mathcal{R}_{EB}$ and $r^{TE}$ decrease forming a roundish cluster centered on $(\mathcal{R}_{EB},\,r^{TE}) \approx (1,\,0.1)$. As structures' major axes become perpendicular to the field, another smaller tail appears with negative $r^{TE}$ and large $\mathcal{R}_{EB}$ values.
This general picture is also observed for sheet-like clouds (Fig.~\ref{fig:REB-rTE_vs_omega-s-f}, bottom right panel) although the scatter of the data points is larger; as expected due to the increase of the degree of freedom.
The comparison between the 2D distributions of $(\mathcal{R}_{EB},\,r^{TE})$ pairs obtained from filament- and sheet-like clouds is further developed below.

\subsection{$\mathcal{R}_{EB}$ and $r^{TE}$ dependence on $\alpha$}\label{alphadep}
Here we infer the dependence of the measured $\mathcal{R}_{EB}$ and $r^{TE}$ as a function of the angle between the magnetic field lines and the line of sight.

\begin{figure*}
    \centering
    $\alpha$ dependence of $\mathcal{R}_{EB}$ and $r^{TE}$\\[-.3ex]
    \includegraphics[trim={2.2cm 4.1cm .4cm 1.cm},clip,width=.8\textwidth]{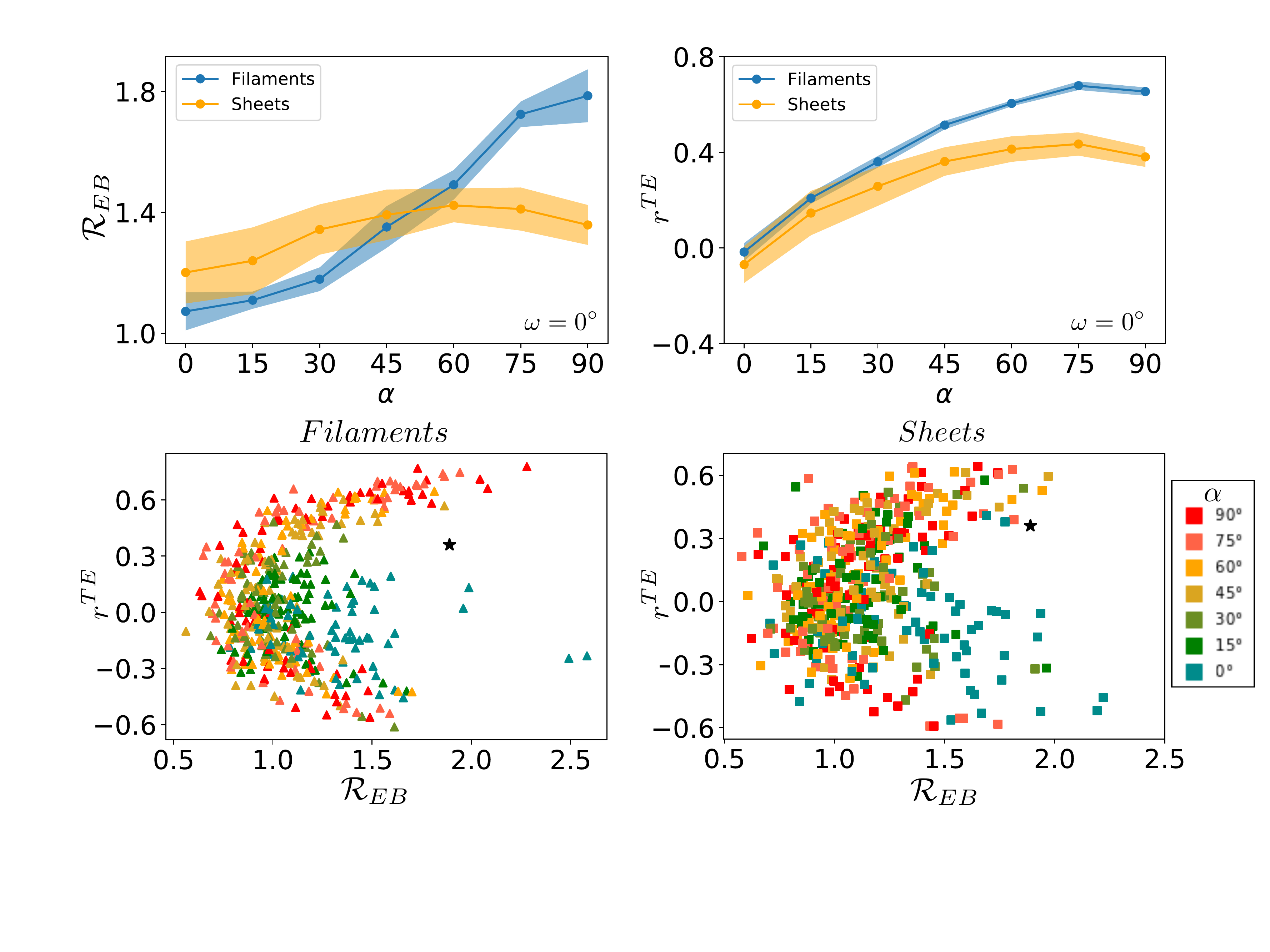}
    \caption{Dependence of $\mathcal{R}_{EB}$ and $r^{TE}$ values on $\alpha$, the angle that makes the magnetic field lines with the line of sight. Top row shows $\mathcal{R}_{EB}$ (left) and $r^{TE}$ (right) versus $\alpha$ for $\omega = 0^\circ$ for both filaments (blue) and sheets (orange). Plotting conventions are the same as in Fig.~\ref{fig:REB-rTE_vs_omega-s-f} except that the color-coded in the bottom row is according to $\alpha$.}
    \label{fig:REB-rTE_vs_alpha-s-f}
\end{figure*}

We first select the synthetic maps created with all the clouds having their major axis parallel to the magnetic field ($\omega = 0^\circ$) and we sort them according to the inclination angle of the field with respect to the line of sight. As in the previous subsection, we have for each $\alpha$ value 10 realizations with filament-like and sheet-like clouds of the same ISM volume that is observed from seven different observer positions (see Sect.~\ref{sec:setup}). A different set of polarization power spectra corresponds to each realization. In the top row of Fig.~\ref{fig:REB-rTE_vs_alpha-s-f}, we show the dependence of the corresponding values of $\mathcal{R}_{EB}$ (left) and $r^{TE}$ (right) as a function of $\alpha$.

For filament-like clouds, the viewing-angle and projection effects on $\mathcal{R}_{EB}$ and $r^{TE}$ are clearly shown. $\mathcal{R}_{EB}$ decreases following a squared-sine trend from its maximum (at $\approx 1.8$) obtained for lines of sight perpendicular to the field lines ($\alpha=90^\circ$) to its minimum (at $\approx 1$) obtained for lines of sight parallel to the field ($\alpha=0^\circ$). As for $r^{TE}$, it also decreases from its maximum when $\alpha \approx 90^\circ$ to zero when $\alpha \approx 0^\circ$ but following a sine trend. For lines of sight and field lines nearly parallel, the degree of polarization is low and any small variations generate polarization with random orientations. This generates equal amounts of $E$ and $B$ modes (thus $\mathcal{R}_{EB} \approx 1$) with no particular correlation with intensity (thus $r^{TE} \approx 0$). 

Similar trends are observed for sheet-like clouds with, however, much reduced difference between extrema. As before, we interpret this reduction by the presence of wiggles and the extra freedom in orientation that sheet-like structures allow for. The maxima of both $\mathcal{R}_{EB}$ and $r^{TE}$ also do not appear for $\alpha = 90^\circ$ but at lower values ($\gtrsim 60^\circ$). We speculate that this peculiarity may be related to projection effects of the 3D structures. 

In the bottom row of Fig.~\ref{fig:REB-rTE_vs_alpha-s-f}, we show the same scatter plots as shown in the bottom row of Fig.~\ref{fig:REB-rTE_vs_omega-s-f} but color-coded by their corresponding inclination angle values.
We observe that the data points corresponding to $\alpha = 90^\circ$ draw a croissant-like pattern (already outlined in the bottom panels of Fig.~\ref{fig:REB-rTE_vs_omega-s-f}) in which $\mathcal{R}_{EB}$ is maximized for $\omega = 0^\circ$ and $90^\circ$ and minimized for $\omega = 45^\circ$, and that is tighter for filament-like clouds than for sheet-like clouds. Then, as $\alpha$ decreases, the distributions of $(\mathcal{R}_{EB},\, r^{TE})$ pairs depart from the croissant shape and become rounder with a center at about $(1,\, 0.1)$. The extreme cases are for $\alpha = 0^\circ$ for which the data points (all $\omega$ values included) show a very large scatter for both filament-like and sheet-like clouds.

\subsection{Ambiguity from the shapes of dust clouds}
It is obvious that sheet-like structures may appear filament-like if they are seen nearly edge-on, that is, if their minor axes is close to the  plane of the sky. Such an example is shown in Fig.~\ref{fig:I_map_ambiguity}. However, if dust clouds of the ISM are sheet-like rather than filament-like structures, and given that generally the 3D orientation of structures does not depend on the observer position (us), there is no reason why all the minor axes of ISM clouds would lie in the plane of the sky%
\footnote{We note that even in the scenario where the Sun is embedded in the Local Bubble and the magnetic field lines follow the surface of the shell of the Local Bubble (as discussed in Sect.~{\ref{sec:discussion}}), it is difficult to find a physical argument that would make the minor axes of all the sheets to lie in the plane of the sky.}. As a consequence, due to projection effects, ripples in the sheets may produce filament-like patterns (ridges) uncorrelated with the projected orientation of the principal axes of the sheets, or of the magnetic field. Therefore, this could induce different patterns in the polarization maps and therefore, different characteristics of the (polarization) power spectra.

We studied the dependence of the summarizing characteristics of the polarization power spectra on the offset angle and on the inclination (viewing) angle for outputs generated out of both, filament-like and sheet-like clouds. Beside the overall similar trend observed for both shape types, we showed that sheet-like structures lead to a generally looser correlation of the $\mathcal{R}_{EB}$ and $r^{TE}$ values than filaments. This reflects the increase of degree of freedom in the projected orientation of substructures in sheets than in filaments. In addition, reaching simultaneously large $\mathcal{R}_{EB}$ and large positive $r^{TE}$ values for sheets appears less common. This increase of the possible ways to project a sheet on the sky results in the reduction of the $E$ to $B$ power asymmetry and the correlation between $T$ and $E$.

\begin{figure}
    \centering
    \begin{tabular}{rc}
      \rotatebox{90}{{\large \hspace{2.3cm} $r^{TE}$}}
            & \includegraphics[trim={1.3cm 1cm 0cm 0cm},clip,width=.9\columnwidth]{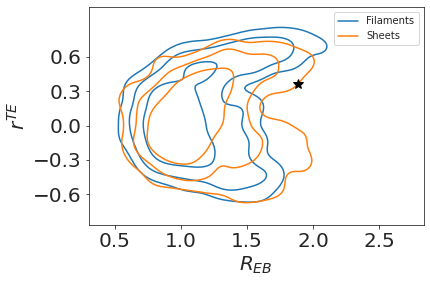} \\
         & {\large  \hspace{.7cm} $\mathcal{R}_{EB}$}
    \end{tabular}
    \caption{Contour plots of the 2D distributions of $(\mathcal{R}_{EB},\,r^{TE})$ from filament-like (blue) and sheet-like (orange) clouds. The embedded contours enclose about 10, 70 and 90 per cent of the data points. To guide the eye, the black star indicates the fiducial Planck values $(\mathcal{R}_{EB}=1.89,\,r^{TE}=0.36)$.}
    \label{fig:contour}
\end{figure}

However, despite these small differences, we find that the distributions of $(\mathcal{R}_{EB},\, r^{TE})$ pairs from maps of filament-like and sheet-like clouds largely overlap. This is best seen in Fig.~\ref{fig:contour} where we show contour plots of the distributions of points on the $(\mathcal{R}_{EB},\, r^{TE})$ plane for both shape categories on a same figure. The three embedded contours enclose about 10, 70 and 90 per cent of the data points.
The overlap is striking and the larger scatter of data points from outputs made out of sheet-like clouds is evident.

Figure~\ref{fig:contour} makes it clear that the shape of the cloud structures cannot be distinguished by the unique consideration of the $\mathcal{R}_{EB}$ and $r^{TE}$ values. The observation of an $(\mathcal{R}_{EB},\, r^{TE})$ pair appears almost as likely coming from either family of shapes. This is even more true if one relaxes the constraints of the morphology (axis ratios) and size of the clouds, which we have kept fixed in this study (see Sect.~\ref{sec:setup}).

\begin{figure}
    \hspace{1.cm} $Filaments$  \hspace{3cm} $Sheets$   \hspace{3cm}\\
    \centering
    \includegraphics[trim={.2cm 1.2cm .3cm 0cm},clip,height=5.1cm]{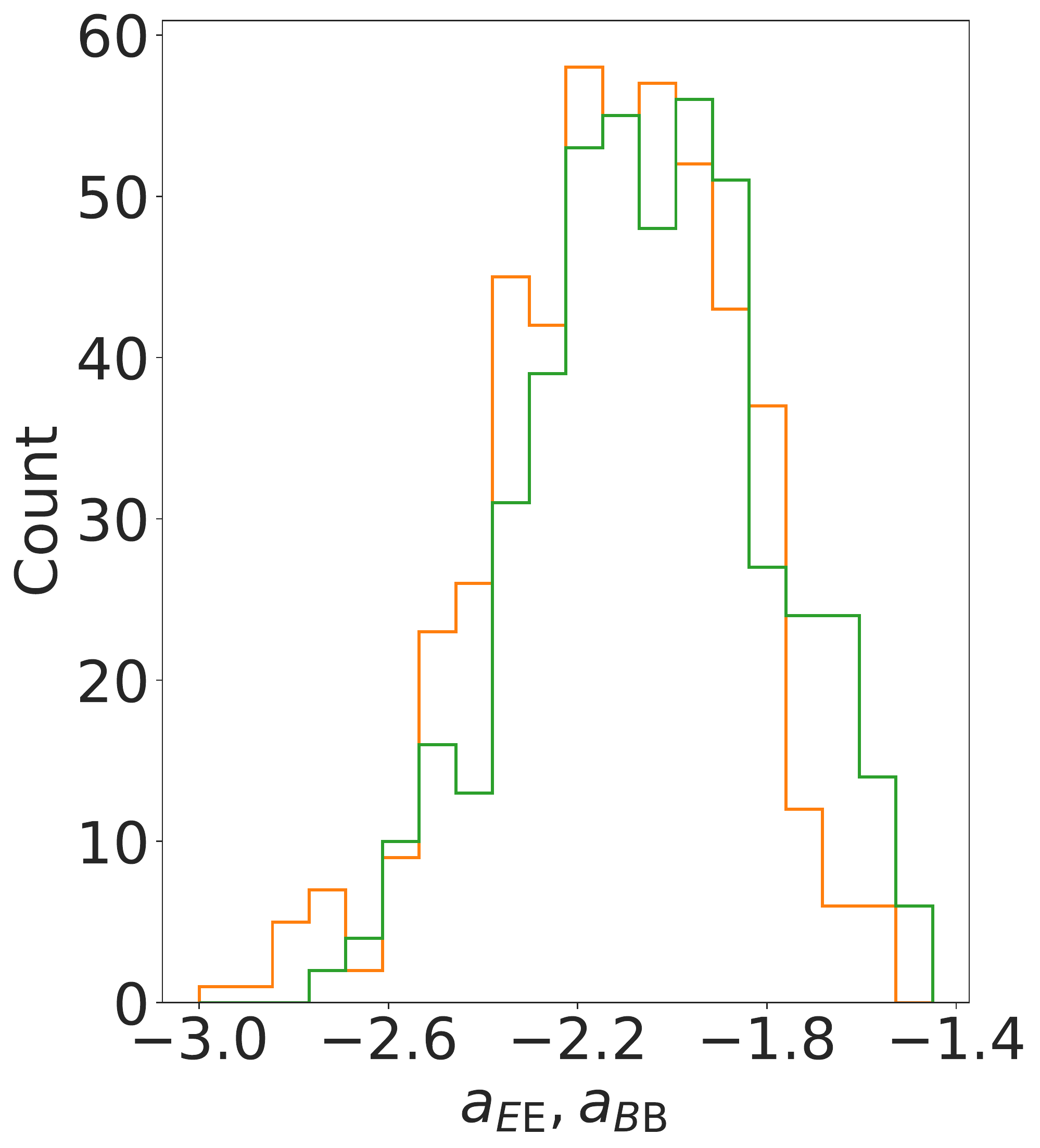}
        \includegraphics[trim={2.9cm 1.2cm .3cm 0.3cm},clip,height=5.1cm]{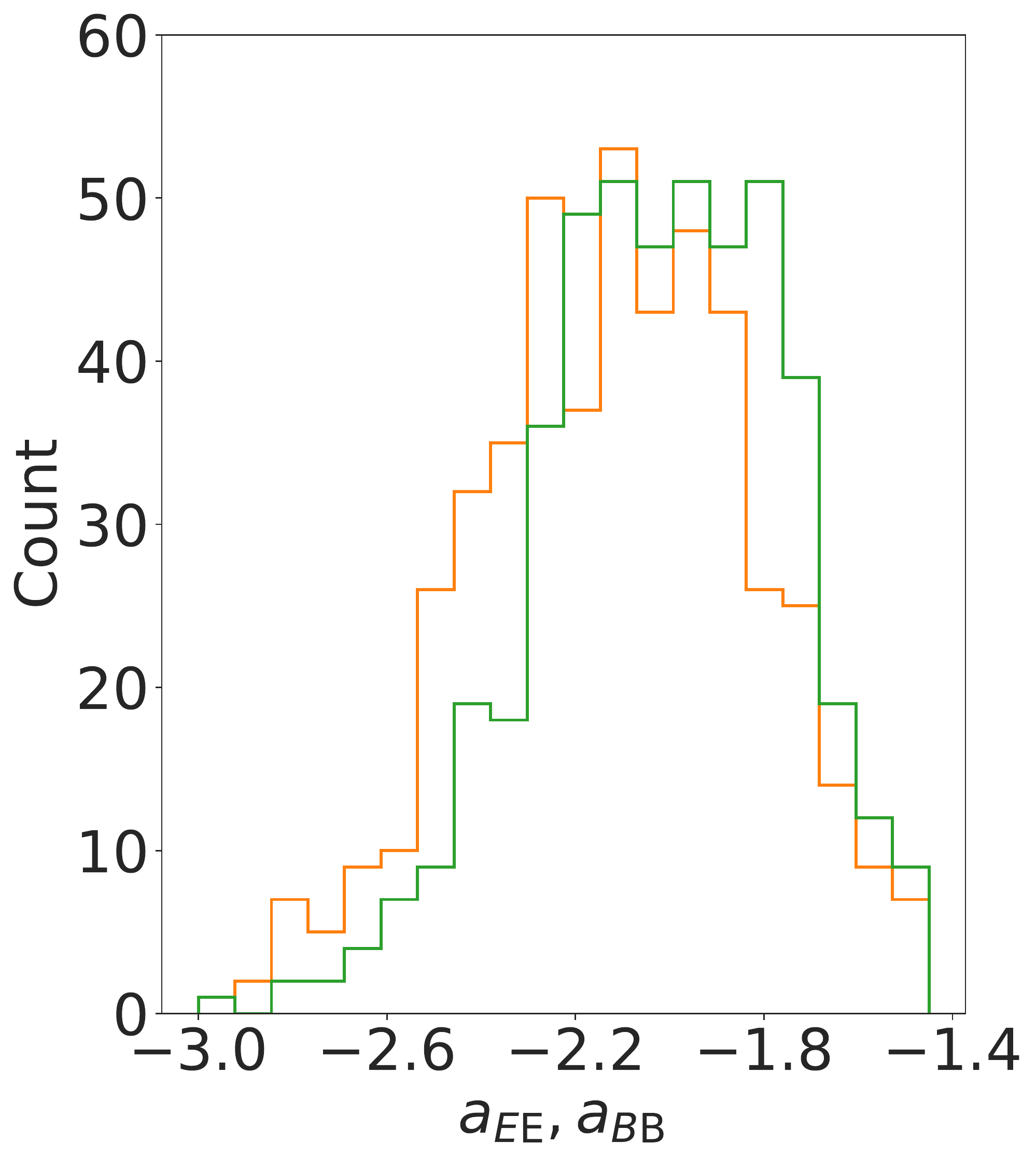}\\[-.8ex]
        \hspace{0.6cm} $a_{EE},\,a_{BB}$  \hspace{2.7cm} $a_{EE},\,a_{BB}$   \hspace{2cm}\\
    \caption{Histograms of spectral indices of the power-law power spectra fitted to the $E$ (orange) and $B$ (green) auto-power spectra obtained for filament-like clouds (left) and sheet-like clouds (right); the full sample is included.}
    \label{fig:spectral_indices}
\end{figure}
Aside from the $\mathcal{R}_{EB}$ and $r^{TE}$ characteristics, power spectra of outputs made from filaments and sheet-like structures are similar. This can already be inferred from the right column of Fig.~\ref{fig:I_map_ambiguity} where, for those examples, it is seen that amplitudes and slopes of spectra have similar values. This is better demonstrated in Fig.~\ref{fig:spectral_indices} where we present histograms of the spectral indices of power-law fits to the auto-power spectrum of $E$ and $B$ modes, for filament (left) and sheet-like (right) clouds. No substantial difference can be spotted between cloud shapes.

\smallskip

Based on these considerations, we argue that interpreting polarization power spectra solely in terms of filament characteristics and their relationship to the ambient magnetic field, is risky because of this degeneracy on the intrinsic shape of the clouds.

\subsection{ Turbulent ISM and magnetic field}
\label{sec:withturbulence}

The toy-model simulations of the magnetized ISM corresponding to the polarization maps used in the analysis so far are somewhat unrealistic in the sense that they lack known physics. In our model, the orientations of density caustics and magnetic field are not correlated at small scales although we expect them to be, at least to some extent, due to flux-freezing (\citealt{Heiles2005}).
These simple, but non-trivial simulations have allowed us to study in detail the dependence of the summary characteristics of the polarization power spectra on the inclination angle, the offset angle and the type of dust cloud morphology.
Here, we want to study the effects of coupling density structures to a non-regular magnetic field on our results. To do so, we exploit the possibility in \textsc{Asterion} to add a stochastic component to the regular magnetic field and to generate density perturbations along the local (perturbed) magnetic field lines (see Sect.~\ref{sec:turbulence}). Example of outputs obtained with such coupling were presented in Figs.~\ref{fig:filshe_Bturbparams} and~\ref{fig:fsh_wiggleparams_v42} for Stokes $Q$ maps  and $K$ maps, respectively.
For the purpose of our analysis we fix the parameters for the stochastic component in magnetic field and for the coupled wiggles such that (i) the outputs are visually similar to the dust polarization sky at high-Galactic latitudes and (ii) the characteristics of polarization power spectra of maps made out of filaments are close to observed values when $\omega = 0^\circ$ and $\alpha = 90^\circ$. This choice, leads to the set of parameters $(n_i,\,n_c) = (0.05,\,0.0015)$ and $(w_i,\,w_c) = (0.2,\,0.001)$ which we do not change while exploring the effect from varying the $\omega$ and $\alpha$ angles as in Sect.~\ref{sec:analysis}.

\begin{figure}
    \centering
    \begin{tabular}{rc}
      \rotatebox{90}{{\large \hspace{2.3cm} $r^{TE}$}}
            & \includegraphics[trim={1.3cm 1cm 0cm 0cm},clip,width=.9\columnwidth]{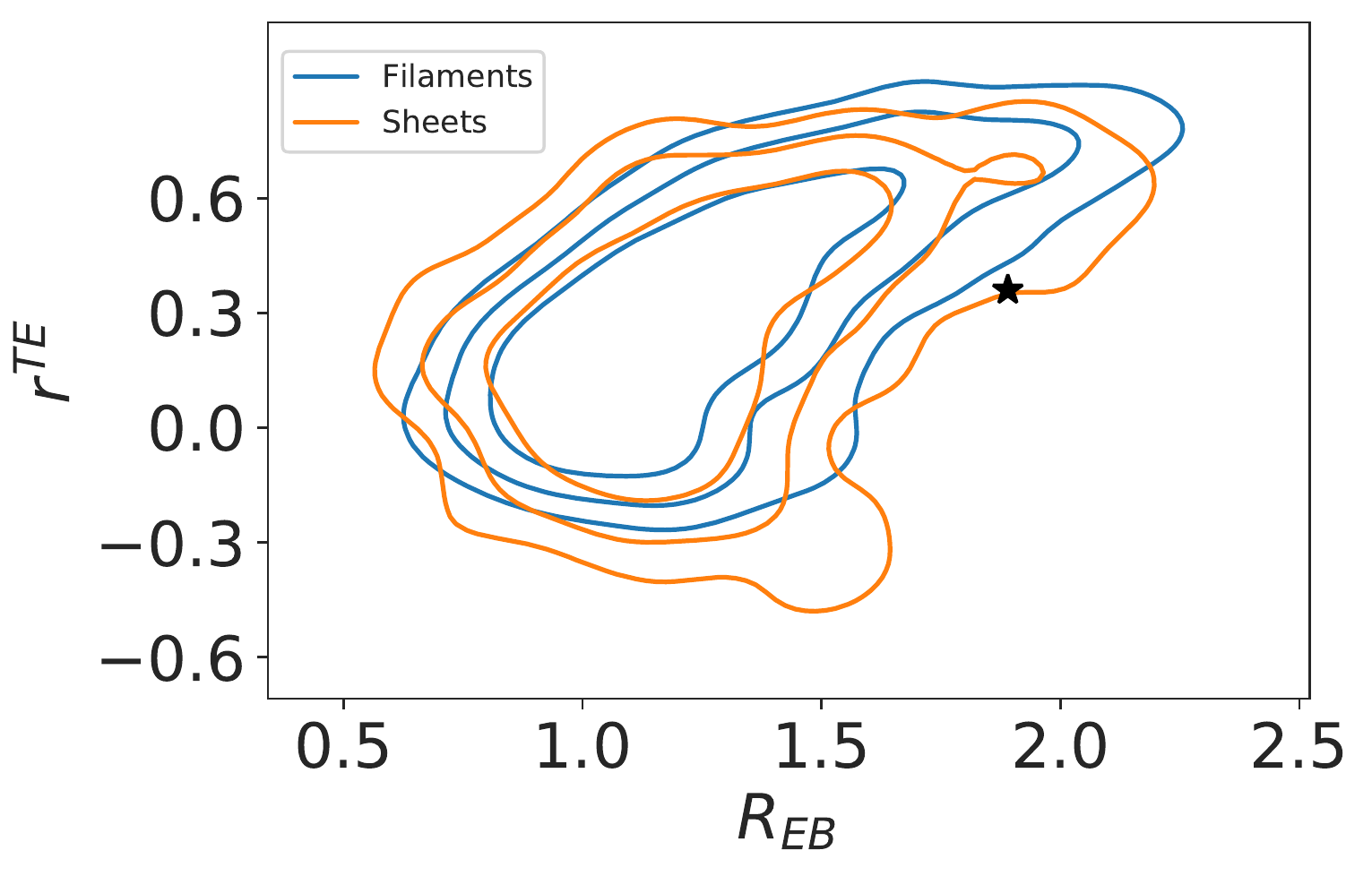} \\
         & {\large  \hspace{.7cm} $\mathcal{R}_{EB}$}
    \end{tabular}
    \caption{
    Same as Fig.~\ref{fig:contour} but with noisy magnetic field and correlated perturbations in density structures.}
    \label{fig:contour_Bnoise}
\end{figure}

With those values we notice that generally the simulated density structures coupled to the magnetic field show their effective major axes more aligned with the local mean magnetic field ($\omega_{\mathrm{eff}} \lesssim 45^\circ$) than set by the input $\omega$ values.
This is a desirable feature motivated by observational evidences that clouds of the diffuse ISM, both H$_{\rm{I}}$ fibers and dust clouds, tend to be preferentially aligned with the ambient magnetic field (\citealt{McC2006};
\citealt{clark14};
\citealt{clark15};
\citealt{mar2015};
\citealt{planck2016XXXII};
\citealt{Planck2016XXXV};
\citealt{planck2016XXXVIII}; \citealt{soler19}).
As a result, and as illustrated in Fig.~\ref{fig:REB-rTE_vs_omega-s-f_Bnoise}, we find that $\mathcal{R}_{EB}$ remains globally larger than unity and $r^{TE}$ remains positive while varying $\omega$. This is the case for both cloud shapes.
Comparing Figs.~\ref{fig:REB-rTE_vs_omega-s-f_Bnoise} and~\ref{fig:REB-rTE_vs_alpha-s-f_Bnoise} with Figs.~\ref{fig:REB-rTE_vs_omega-s-f} and~\ref{fig:REB-rTE_vs_alpha-s-f}, we find that the different scalings present for the range $\omega \gtrsim 30^\circ$ are severely attenuated, if not suppressed, when density structures are coupled to the magnetic field and that $\mathcal{R}_{EB}$ and $r^{TE}$ values of outputs from sheet-like structures are in general closer to outputs from filament-like structures. The main effects of coupling the density perturbations and the magnetic field is that 2D distributions of $\mathcal{R}_{EB}$ and $r^{TE}$ from the two types of shapes do overlap even more than in the absence of such coupling and that negative $r^{TE}$ values are much less frequent. This is clearly seen in Fig.~\ref{fig:contour_Bnoise}.

\section{Discussion}
\label{sec:discussion}

\smallskip

In this paper we do not attempt to model the actual dust polarization sky observed at submillimeter wavelengths, for example by Planck at 353~GHz. Rather, our focus is to explore whether the assumed shape of the ISM clouds may lead to erroneous conclusions in the interpretation of dust polarization power spectra. 

In order to model the dust polarization sky and in particular the measured polarization power spectra, it would be necessary to sample a large volume of the Galactic space with our OB, generating a 3D mosaic, possibly in a way similar to what has been proposed by \cite{her2021}. However, our analysis shows that the viewing angle -with which the portion of the ISM is observed- is significant in determining the ratio and correlation between power spectrum modes. We find an extra dependence on $\alpha$ than that proposed by \cite{huffenberger} for $r^{TE}$ only. Our results make it clear that the overall geometrical arrangement of the magnetized ISM surrounding the observer (us) leaves its marks on the power spectra and in particular on the $\mathcal{R}_{EB}$ and $r^{TE}$ parameters. We see it unlikely that the geometry dependence averages out when considering large swaths of the sky. This is reinforced by the fact that all viewing angles are not equally likely, especially at high Galactic latitudes. The reason is that the Sun sits in the Local Bubble, a cavity of hot plasma that was presumably created by supernova explosions, which is surrounded by a shell of cold neutral gas and dust with a radius between 100 to 300 pc, depending on observation direction. The polarized signal observed in the polar caps ($|b|>60^\circ$) is dominated by the emission from the shell of the Local Bubble (\citealt{Ska2019}). The formation of this structure has shaped the local, magnetized ISM surrounding the Sun (\citealt{Alv2018}; \citealt{pel20}) and has left the magnetic field to lie mostly in the plane of the sky at high $|b|$ but may show strong departures in some places (\citealt{pel20}). In light of the highlighted $\alpha$-dependence that we find, we caution that the exact morphology of the magnetic field in regions dominating the polarization signal needs to be accounted for to model large areas of the sky. This would involve the production of 3D mosaics, and the computation of polarization power spectra to be compared to actual observations. This is a highly complex task, which is beyond the scope of this paper.

Recently, \cite{PelNT22} showed from maps synthesized from Galaxy-sized MHD simulations that the specific observer location has a strong effect on the characterization of the polarization power spectra. In light of our results, it would be interesting to figure out how much of the reported variance can be attributed to the peculiar 3D geometry of the part of the magnetic field that is imprinted in maps and thus, in power spectra.

On the other hand, \cite{bracco} have studied polarization power spectra from Planck data at intermediate-to-low Galactic latitudes in sky patches of the same size as ours. They found that the $\mathcal{R}_{EB}$ and $r^{TE}$ parameters span large ranges of values. With our toy-model simulation-based study we show that a significant source of scatter in those quantities may actually come from the geometrical arrangement of the magnetized ISM, in addition to the scatter that may arise from the fact that the offset angle in different clouds is probably not constant but rather follows some distribution (see e.g., \citealt{huffenberger}).
Our toy-models, however, cannot be simply compared to the results of their observational study since our synthetic maps do not account for the full cone of observation, which becomes important at low latitudes, and also because they pre-processed the polarization sky maps in order to get rid of the large-scale Galactic gradient. We do not know how this processing affects their results.

Furthermore, in an attempt to model the actual sky, it would probably be relevant to consider that dust clouds can be described by a mixed population of sheet-like and filament-like structures with varying axis ratios.
In this study, we did not explore such a possibility because we want to quantify the possible cloud-shape degeneracy of the $\mathcal{R}_{EB}$ and $r^{TE}$ values.

Currently, our results are limited by the achieved spatial resolution that is fixed by the minimum angular size of the sky patch that \texttt{Xpol} can handle, and the size of \textsc{Asterion} grid ($256^3$). We achieve a spatial resolution of $\sim 0.78$ pc which is more than twice the thickness obtained for actual sheet clouds of the neutral ISM ($\lesssim 0.3$ pc according to \citealt{kalberla}).

Another source of limitation of our study comes from the limited portion along distance that is mapped. This reduces the number of structures that appear on our maps as compared to real observations. This does not affect our analysis since our main results relate to the (relative) comparison of power spectrum characteristics between (i) different cloud shapes, (ii) different offset angles, and (iii) different inclination angles.

\medskip

Besides the special attention paid to the $E/B$ power asymmetry and the correlation between $T$ and $E$ modes which were detected with high significance in dust polarization maps from Planck, a more marginal signal has been reported between the $T$ and $B$ modes (\citealt{planck16XXX}; \citealt{planck20XI}). Other recent studies (\citealt{wei2020}; \citealt{Cla+21}) relied on a more accurate version of the Planck data and on external data sets to prove that the weak positive $TB$ signal (a mark of parity violation in the dust polarization sky) is a real property of the Galactic emission and infer its origin.

A detailed study of the effect of cloud morphology on the $TB$ signal (e.g., Bracco et al. 2019a) and a careful analysis of the dependence on $\omega$ and $\alpha$ along with the scale dependence of the signal is not the focus of this paper and we postpone this to future work. Meanwhile, we show in Fig.~\ref{fig:rTB}, that the correlation coefficient $r^{TB}$ between $T$ and $B$ modes measured from filament-like or sheet-like structures does not strongly differ, at least when the scatter from varying $\omega$ and or $\alpha$ is taken into account. All distributions are centered on zero and, as with $\mathcal{R}_{EB}$ and $r^{TE}$, we find that sheet-like structures lead to broader distributions than filament-like structures do.

{\begin{figure}
    \hspace{1.1cm} Regular $\textbf{B}$ field  \hspace{1.4cm} Turbulent ISM \hspace{3cm}\\
    \centering

    \includegraphics[trim={.2cm 1.6cm .2cm 0cm},clip,height=5.1cm]{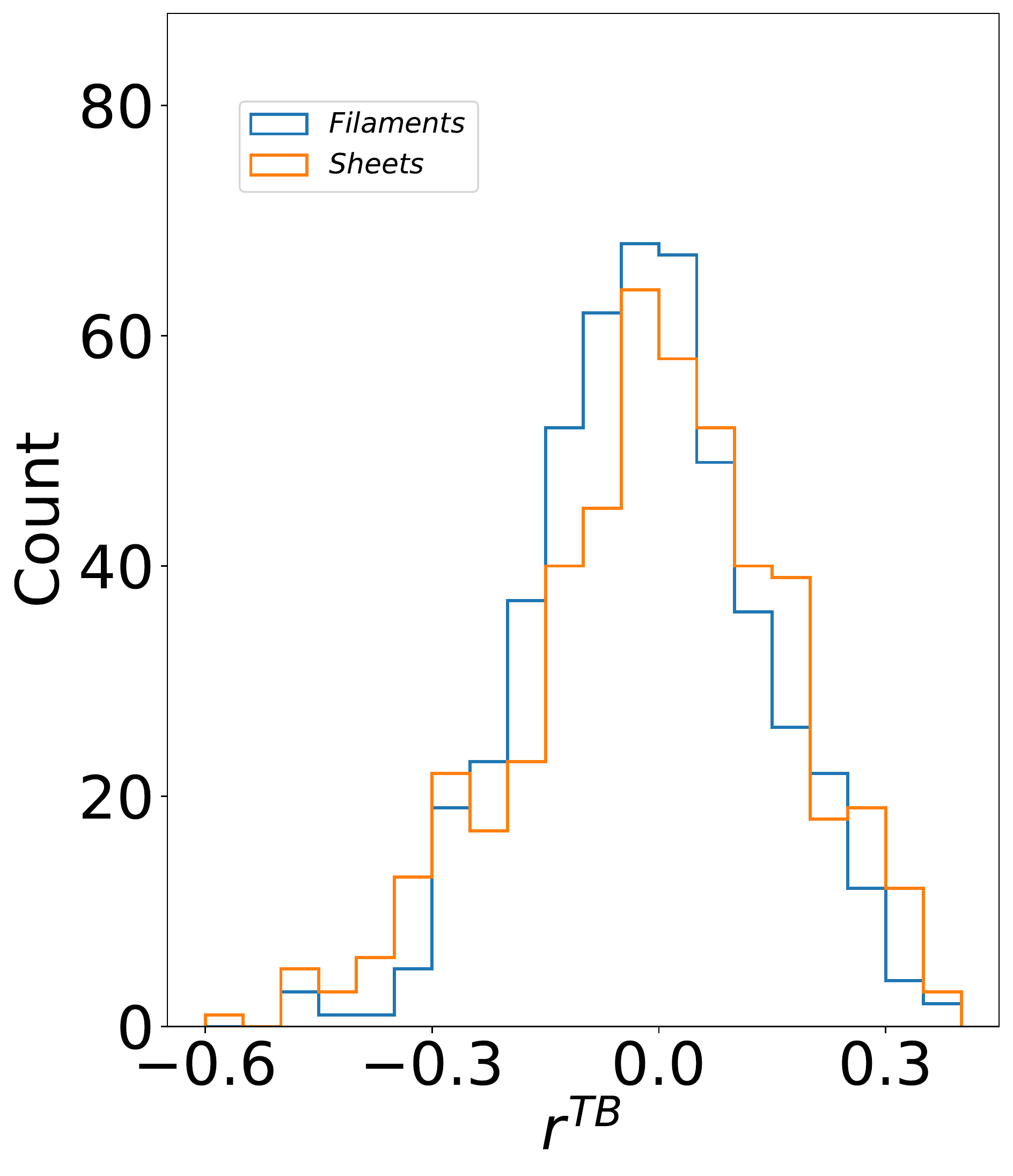}
        \includegraphics[trim={2.9cm 1.6cm .2cm 0cm},clip,height=5.1cm]{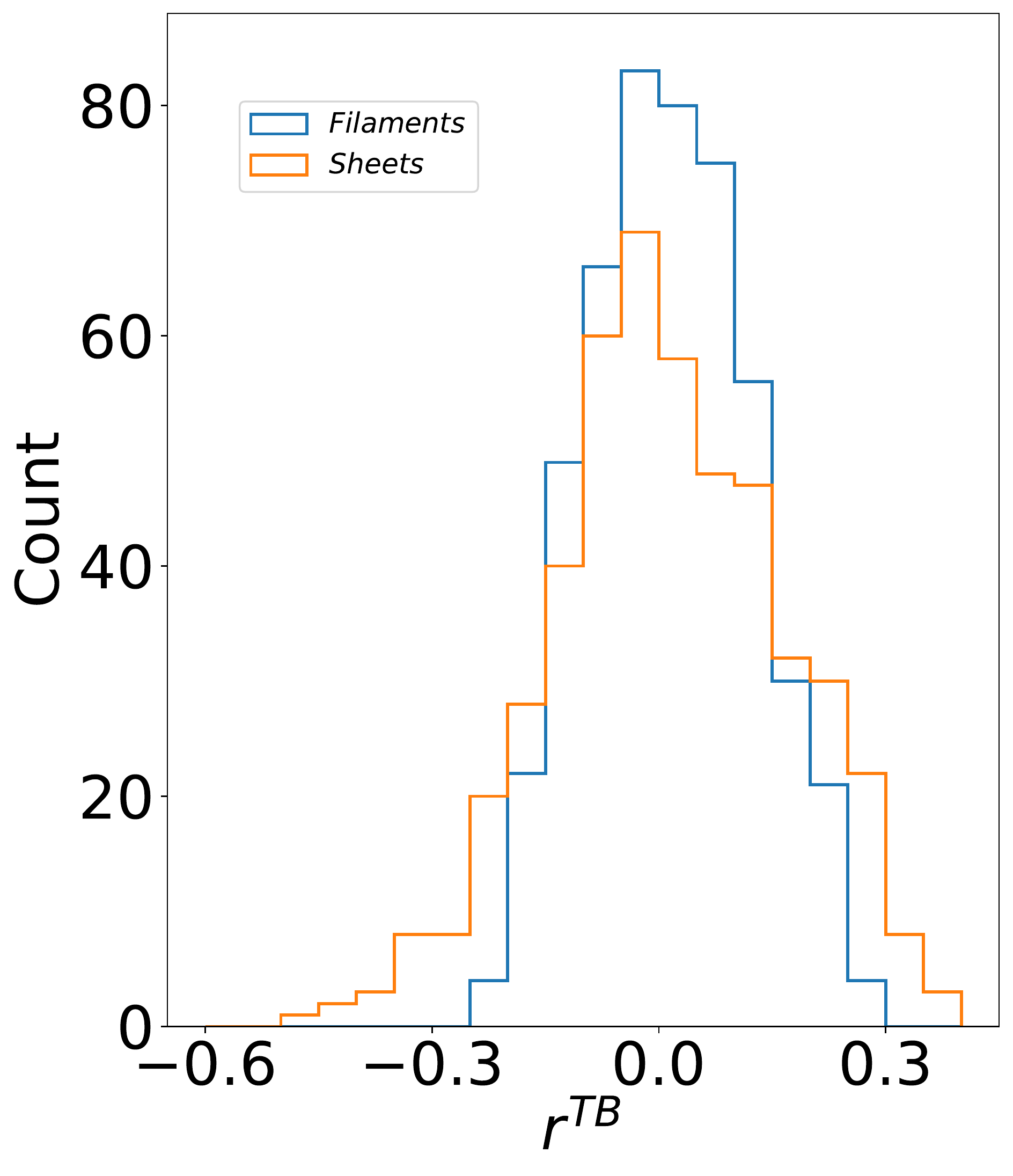}\\[.5ex]
    \hspace{1.15cm} {$r^{TB}$}   \hspace{3.5cm} {$r^{TB}$}   \hspace{3cm} \\[-.5ex]
    \caption{Histograms of $r^{TB}$ obtained for filament-like clouds (blue) and sheet-like clouds (orange) for { regular magnetic field (left) and with "turbulence" switched on} (right); the full sample is included.}
    \label{fig:rTB}
\end{figure}
}

\section{Conclusions}
\label{sec:conclusion}

The search for the primordial $B$ modes in the polarization of CMB radiation calls for the understanding and characterization of the Galactic foregrounds in order to disentangle their contribution to the maps of the polarized sky.

In this project, we investigated whether the morphology of the dust clouds has an impact on the statistical characterization of the polarized emission. We simulated interstellar clouds with filament-like and sheet-like shapes and produced corresponding synthetic polarization maps, using the software \textsc{Asterion}. 
Then, we computed their polarization power spectra in multipole range $\ell \in [100,\, 500]$ and we focused our study on the power asymmetry between $E$ and $B$ modes, and on the cross correlation between $T$ and $E$ modes, through the parameters $\mathcal{R}_{EB}$ and $r^{TE}$, respectively. We explored the dependence of $\mathcal{R}_{EB}$ and $r^{TE}$ on the inclination angle ($\alpha$, between magnetic field and the line of sight), and the offset angle ($\omega$, between the longest cloud axis and the magnetic field) for both types of shape.
Our study provides important insights on the interpretation of the polarization power spectra of dust polarization in terms of ISM properties and could help in the characterization of Galactic foregrounds to the CMB polarization.

\smallskip

Our results for filaments are consistent with theoretical predictions and observations. 
For filamentary structures aligned with the magnetic field ($\omega \approx 0^\circ$), something that seems to be true in nature for the diffuse ISM, the power ratio $\mathcal{R}_{EB}$ can be as large as $\sim$ 2, in agreement with Planck measurements (e.g., \citealt{planck16XXX}). Moreover, for regular 
magnetic field and large inclination angles, 
where polarization is maximized, $E$-mode power of structures is maximized for offset angles 
$\omega$ = 0$^\circ$ and 90$^\circ$ and $B$-mode power is maximized for $\omega$ = 45$^\circ$.
This behaviour is in agreement with theoretical expectations \cite{zaldarriaga} and previous studies (\citealt{rot2019}; \citealt{huffenberger}).
As a result, $\mathcal{R}_{EB}$ as a function of $\omega$ draws a parabola with minimum at $\omega \approx 45^\circ$, where $B$ modes slightly dominate $E$ modes.
In parallel, the correlation coefficient $r^{TE}$ goes from its maximum at $\omega=0^\circ$ to its minimum at $\omega = 90^\circ$.
These trends (with little variance from specific ISM realizations) are similarly observed for other inclination angle values, though with lower difference between extrema as $\alpha$ decreases.

To the best of our knowledge, this is the first time that an analysis of polarization power spectra obtained from sheet-like interstellar clouds has been performed. We found that the parameters $\mathcal{R}_{EB}$ and $r^{TE}$ measured from maps made from sheet-like structures show similar trends as a function of $\omega$ and $\alpha$ angles to those observed for filaments. However, the amplitude between extrema is generally smaller and the variance with respect to the specific ISM realizations is larger.
We understood these characteristics in terms of the extra degree of freedom that sheet-like structures possess related to the orientation of their corrugations with the long axis. 

\smallskip

Varying only the inclination and offset angles, we
found that sheet-like structures and filament-like structures lead to polarization power spectra that cannot be distinguished in the plane of $(\mathcal{R}_{EB},\,r^{TE})$ parameters.

Consequently, we argued that measured $\mathcal{R}_{EB}$ and $r^{TE}$ values cannot be used alone to discriminate among characteristics of ISM structures and that this degeneracy should be accounted for in the interpretation and modeling of observational data.
We checked that our main conclusions remain valid when a stochastic component is added to the large-scale magnetic field and, most importantly, when density structure perturbations are added such that the orientations of density caustics and magnetic field are correlated at small scales as we expect them to be due to MHD physics.

\smallskip

Our analysis also highlights and quantifies the dependence of the $E/B$ asymmetry and $TE$ correlation on the inclination  angle, a dependence that is changed by the inclusion of fluctuations in the magnetic field lines and their coupling to density perturbations but that preserves its significance.
This strongly suggests that the geometrical arrangement of the large scale magnetic field as viewed by the observer is a critical factor shaping polarization power spectra.

This was already suggested by \cite{bra2019} and could explain the large scatter on $\mathcal{R}_{EB}$ and $r^{TE}$ values observed from different sky patches (\citealt{bracco}) and the strong cosmic variance reported from MHD simulation-based studies (\citealt{kim2019}; \citealt{PelNT22}).

Finally, and for completeness, we also examined the correlation coefficient between the $T$ and $B$ modes ($r^{TB}$) and observe that this feature of the power spectrum cannot be used to discriminate cloud shape types either. Instead, we find that variation in $\omega$ and $\alpha$ generates significant dispersion in the values of $r^{TB}$.
Further work will be needed to investigate the possible scale dependence of the $TB$ signal as reported by real sky observation of dust polarization.

\begin{acknowledgements}
We would like to thank G. Panopoulou, R. Skalidis, T. Ghosh, and V. Pavlidou for insightful discussions related to this project.
We warmly thank our referee, Andrea Bracco, for his thorough review and for providing us with sound comments which help us improve the quality of this paper.
This project has received funding from the European Research Council (ERC) under the European Unions Horizon 2020 research and innovation programme under grant agreement No. 771282.
\end{acknowledgements}

\bibliographystyle{aa}
\bibliography{citations}
%

\begin{appendix}

\section{Power spectrum characteristics and turbulent ISM}
{

Figures~\ref{fig:REB-rTE_vs_omega-s-f_Bnoise} and~\ref{fig:REB-rTE_vs_alpha-s-f_Bnoise} show the main results obtained when we consider our simple implementation to account for density structures correlated to non-regular magnetic field, that is, to mimic basic effects expected from MHD physics. These figures are obtained following the very same procedure as  Figs.~\ref{fig:REB-rTE_vs_omega-s-f} and~\ref{fig:REB-rTE_vs_alpha-s-f} obtained with a uniform magnetic field permeating the OB.
\begin{figure*}
    \centering
    $\omega$ dependence of $\mathcal{R}_{EB}$ and $r^{TE}$\\[-.3ex]
    \includegraphics[trim={1.2cm 2.6cm 1.2cm 2.2cm},clip,width=.8\textwidth]{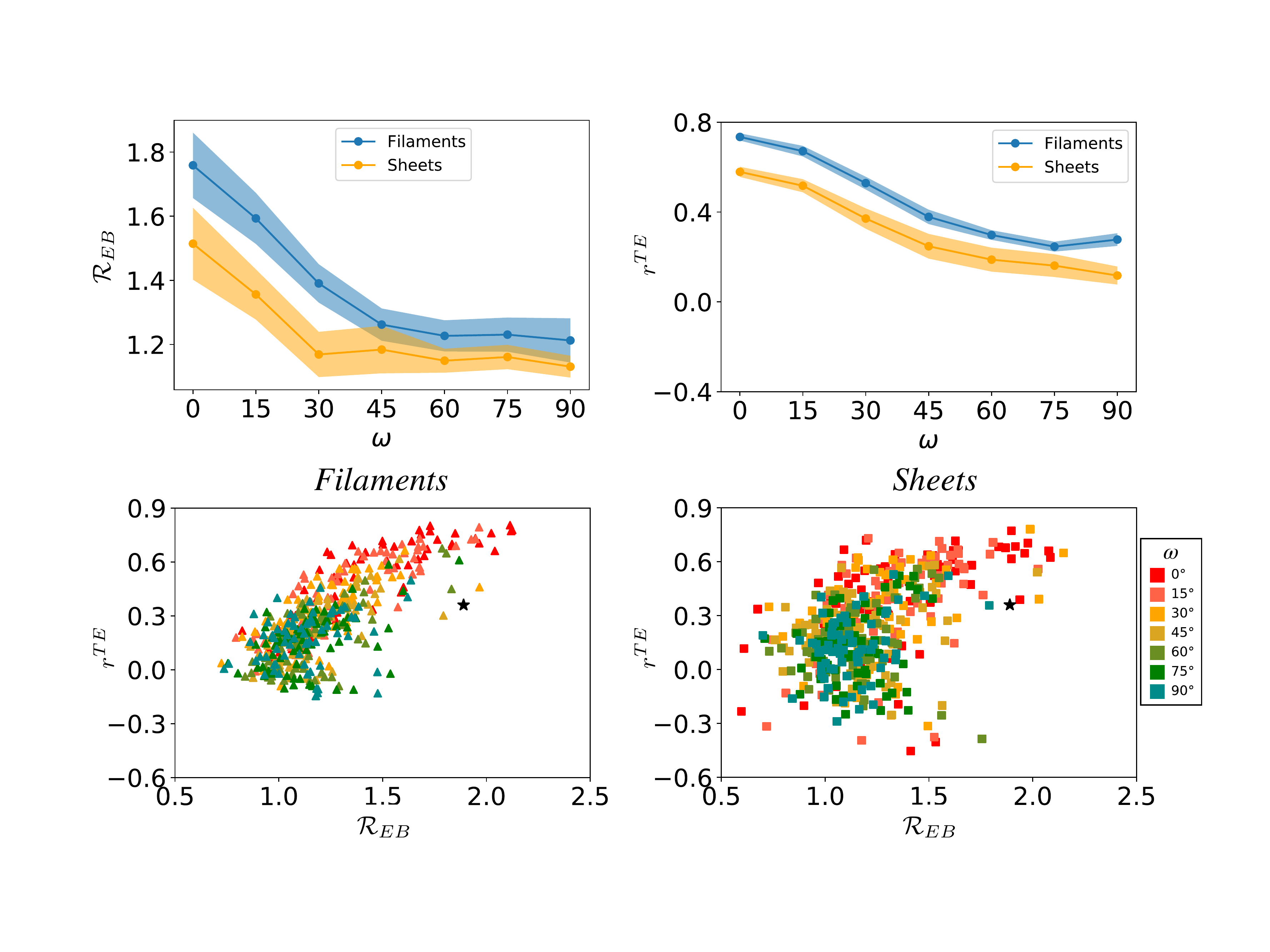}
    \caption{
    Same as for Fig.~\ref{fig:REB-rTE_vs_omega-s-f} but for when a 3D vector noise is added to the regular magnetic field and the wiggles of density structures are constrained to follow the local magnetic field lines.}
    \label{fig:REB-rTE_vs_omega-s-f_Bnoise}
\end{figure*}
\begin{figure*}
    \centering
    $\alpha$ dependence of $\mathcal{R}_{EB}$ and $r^{TE}$\\[-.3ex]
    \includegraphics[trim={1.2cm 2.6cm 1.2cm 2.2cm},clip,width=.8\textwidth]{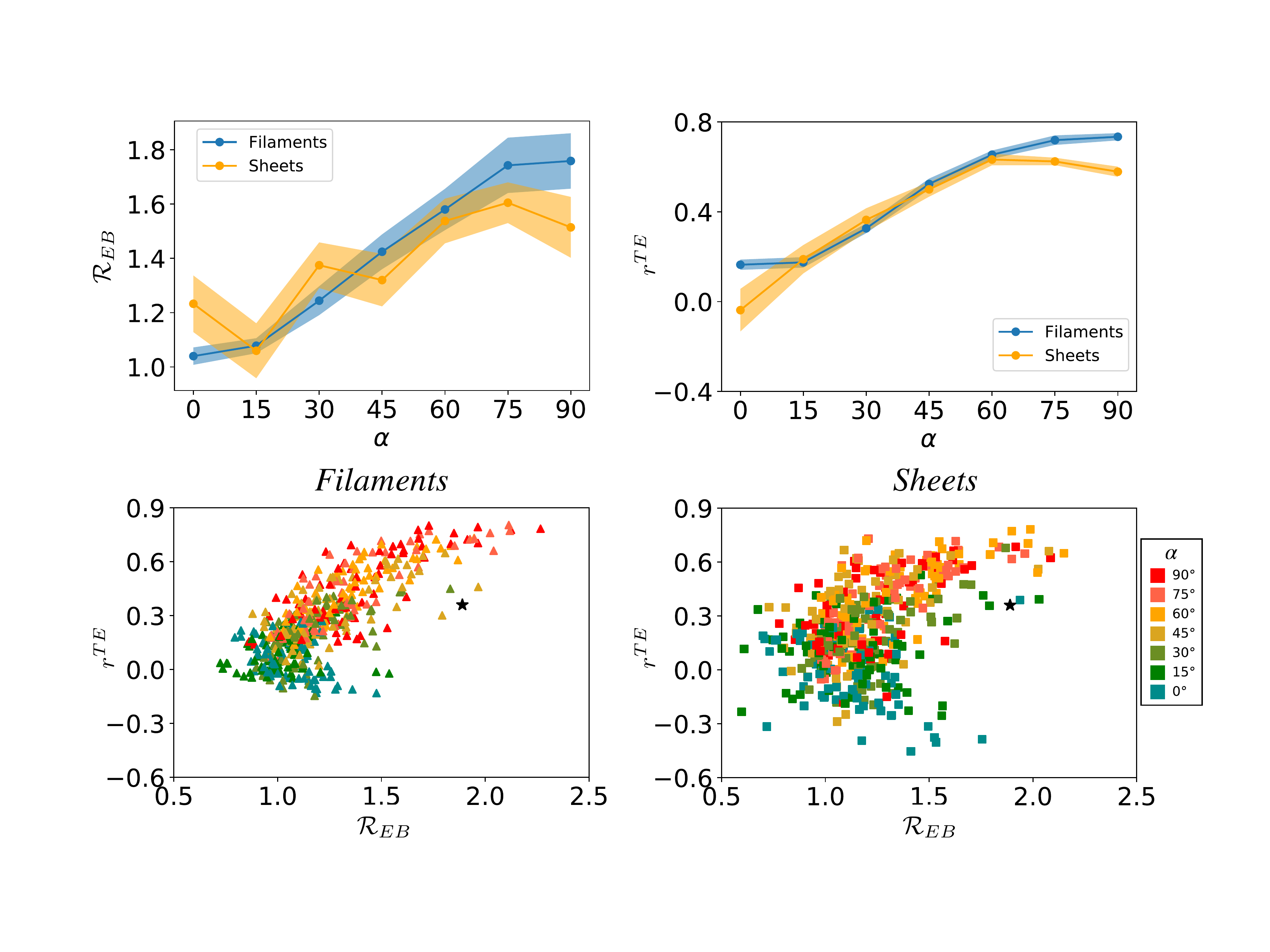}
    \caption{
    Same as for Fig.~\ref{fig:REB-rTE_vs_alpha-s-f} but for when a 3D vector noise is added to the regular magnetic field and the wiggles of density structures are constrained to follow the local magnetic field lines.}
    \label{fig:REB-rTE_vs_alpha-s-f_Bnoise}
\end{figure*}
}

\end{appendix}

\end{document}